%Paper: hep-th/9511053
%From: atish@theory.caltech.edu (Atish Dabholkar)
%Date: Tue, 7 Nov 95 18:13:24 PST

%
%
%-----------------------------
%  This paper uses harvmac
%-----------------------------
\input harvmac.tex
\noblackbox
%
% Definitions
%
\def\half{{1\over2}}

\overfullrule=0pt
\def\Title#1#2{\rightline{#1}\ifx\answ\bigans\nopagenumbers\pageno0\vskip1in
\else\pageno1\vskip.8in\fi \centerline{\titlefont #2}\vskip .5in}

\baselineskip=18pt plus 2pt minus 2pt

\def\ajou#1&#2(#3){\ \sl#1\bf#2\rm(19#3)}
%
%Caligraphic
%

\def\apm{$\alpha '$}
%
%\def\eightbig#1{{\hbox{$\textfont0=\ninerm\textfont2=\niesy\left#1\vbox
% to6.5pt{}
%\right.\n@space$}}}
%\eightbig{N}
%
%Roman

%
%Greek
%

\def\r{\rho}

\def\s{\sigma}

\def\th{\theta}
\def\a{\alpha}
\def\b{\beta}
\def\d{\delta}
\def\m{\mu}
\def\n{\nu}
\def\p{\pi}
\def\e{\epsilon}
\def\w{\omega}

\def\Sig{\Sigma}
\def\D{\Delta}

%
%This paper
%
\font\cmss=cmss10 \font\cmsss=cmss10 at 7pt

\def\av#1{{\langle #1 \rangle}}
\def\tphi{{\tilde \phi}}
\def\vf{\vec F}
\def\vx{\vec x}
\def\fdot{{\dot{\vec F}}}
\def\fd{{\dot F}}
\def\fds{{\dot F}^2}
%
%Shortforms
%
\def\IZ{\relax\ifmmode\mathchoice
{\hbox{\cmss Z\kern-.4em Z}}{\hbox{\cmss Z\kern-.4em Z}}
{\lower.9pt\hbox{\cmsss Z\kern-.4em Z}}
{\lower1.2pt\hbox{\cmsss Z\kern-.4em Z}}\else{\cmss Z\kern-.4em Z}\fi}
\def\IC{\relax\hbox{$\inbar\kern-.3em{\rm C}$}}
\def\IR{\relax{\rm I\kern-.18em R}}
\def\1{\relax 1 { \rm \kern-.35em I}}
\def\frac#1#2{{#1 \over #2}}
\def\ie{{\it i.e.}}

\def\p+{{\partial_+}}

\def\half{{1 \over 2}}

\def\apm{\alpha^{\prime}}
%Old

\def\hf{{1\over2}}

\def\ajou#1&#2(#3){\ \sl#1\bf#2\rm(19#3)}
\def\apm{{\alpha^\prime}}

\def\frac#1#2{{#1 \over #2}}

\def\etp{e^{2\phi}}

\def\mn{{\mu\nu}}

\def\vx{{\vec x}}
\def\vX{{\vec X}}

\def\p{\partial}

\def\nb{\nabla}
%
% reference defs
%
\def\npb#1#2#3{{\sl Nucl. Phys.} {\bf B#1} (#2) #3}
\def\plb#1#2#3{{\sl Phys. Lett.} {\bf #1B} (#2) #3}
\def\prl#1#2#3{{\sl Phys. Rev. Lett. }{\bf #1} (#2) #3}
\def\prd#1#2#3{{\sl Phys. Rev. }{\bf D#1} (#2) #3}

%-------------------
% references
%-------------------
%
\lref\bhd{J. Blum and J. A. Harvey, unpublished; M. J. Duff, J. T. Liu,
and R. Minasian, \npb{452}{1995}{261}, hep-th/9506126.}
\lref\bekal{K.Behrndt, ``About a Class of Exact String Backgrounds",
HUB-EP-9516, hep-th/9506106\semi
R. Kallosh and A. Linde, ``Exactly Supersymmetric Massive and
Massless Black Holes", SU-ITP-95-14, hep-th/9507022.}
\lref\busch{T. H. Buscher, \plb{194}{1987}{59}; \plb{201}{1988}{466}.}
\lref\inflow{C. G. Callan and J. A. Harvey, \npb{250}{1985}{427}.}
\lref\hullwitt{C. Hull and E. Witten, \plb{160}{1985}{398}.}
\lref\schsen{J. Schwarz and A. Sen, \plb{312}{1993}{105},
hep-th/9305185.}
\lref\senentrop{A. Sen, {\sl Mod. Phys. Lett.} {\bf A10} (1995) 2081,
hep-th/9504147.}
\lref\brane{J. Polchinski, ``Dirichlet Branes and Ramond-Ramond
Charges", NSF-ITP-95-122, hep-th/9510017.}
\lref\braneone{J. Polchinski and E. Witten, ``Evidence for Heterotic-Type I
String Duality", IASSNS-HEP-95-81/NSF-ITP-95-135, hep-th/9510169.}
\lref\branetwo{P. Horava and E. Witten,
``Heterotic and Type-I String Dynamics from Eleven Dimensions,''
IASSNS-HEP-95-86/PUPT-1571, hep-th/9510209.}
\lref\wittms{E. Witten, {\sl Phys. Lett.} {\bf B153} (1985) 243. }
\lref\hm{J. A. Harvey and G. Moore, ``Algebras, BPS states, and strings,''
hep-th/9510182.}
\lref\dabhharv{ A. Dabholkar and J. A. Harvey,
\prl{63}{1989}{478}.}
\lref\dghr{ A. Dabholkar, G. Gibbons, J. A. Harvey and F. R. Ruiz-Ruiz,
\hfill\break \npb{340}{1990}{33}.}
\lref\harvstro{ J. A. Harvey and A. Strominger,
\npb{449}{1995}{535}, hep-th/9504047.}
\lref\gsw{M. B. Green, J. H. Schwarz, and E. Witten,
{\it Superstring Theory},  {\rm Vols. I and II},
\hfill\break Cambridge University Press (1987).}
\lref\mtw{C. W. Misner, K. S. Thorne, and J. A. Wheeler,
{\it Gravitation}, W. H. Freedman and Company (1973).}
\lref\ghs{D. Garfinkle, G. Horowitz and A. Strominger,
\prd{43}{1991}{3140}.}
\lref\hands{G. Horowitz and A. Sen,
``Rotating Black Holes which Saturate a Bogomol'nyi Bound," UCSBTH-95-27,
hep-th/9509108.}
\lref\senchsol{A. Sen, \npb{388}{1992}{457}, hep-th/9206016.}
\lref\senetal{A. Sen. \plb{271}{1991}{295}\semi
A. Sen. \plb{274}{1991}{34}\semi
S. Hassan and A. Sen, \npb{375}{1992}{103}.}
\lref\sensix{ A. Sen, \npb{450}{1995}{103},
hep-th/9504027.}
\lref\senblack{A. Sen, \npb{440}{1995}{421}}
\lref\senfour{A. Sen, {\sl Int. J. Mod. Phys.}
{\bf A9}  (1994) 3707,  hep-th/9402002.}
\lref\swa{N. Seiberg and E. Witten, \npb{426}{1994}{19},
hep-th/9407087.}
\lref\swb{N. Seiberg and E. Witten,
\npb{431}{1994}{484}, hep-th/9408099.}
\lref\wittoliv{E. Witten and D. Olive, \plb{78}{1978}{97}.}
\lref\foursusy{C. Montonen and D. Olive, \plb{72}{1977}{117}\semi
P. Goddard, J. Nuyts and D. Olive, \npb{125}{1977}{1}\semi
H. Osborn, \plb{83}{1979}{321}.}
\lref\russsuss{J. Russo and L. Susskind, \npb{437}{1995}{611},
hep-th/9405117.}
\lref\veneziaetal{G. Veneziano, \plb{265}{1991}{287}\semi
K. Meissner and G. Veneziano, \plb{267}{1991}{33}\semi
M. Gasperini, J. Maharana and G. Veneziano, \plb{272}{1991}{277}.}
\lref\azcar{J. A. de Azcarraga, J. P.~Gauntlett, J. M. Izquierdo
and P. K. Townsend, \prl{63}{1989}{2443}.}
\lref\ghl{J. P.~Gauntlett, J. A. Harvey, and J. T.~Liu,
\npb{409}{1993}{363}.}
\lref\khuri{R. Khuri, \npb{387}{1992}{315}.}
\lref\horotsey{G. Horowitz and A. Tseytlin, \prd{50}{94}{5204},
hep-th/9408040; \prd{51}{95}{2896}, hep-th/9409021.}
\lref\wittD{E. Witten, ``Bound States of Strings and P-Branes",
IASSNS-HEP-95-78, hep-th/9510335.}
\lref\wittduality{E. Witten, \npb{443}{1995}{85}, hep-th/9503124.}
\lref\hulltown{C.M. Hull and P.K.  Townsend, \npb{438}{1995}{109},
hep-th/9410167. }
\lref\duffss{M. Duff, \npb{442}{1995}{47}, hep-th/9501030.}
\lref\hl{J. A. Harvey and J. Liu, {\sl Phys. Lett.} {\bf B268} (1991) 40.}
\lref\vachetal{T. Vachaspati, {\sl Nucl. Phys.} {\bf B277} (1986) 593\semi
Vachaspati and T. Vachaspati,
{\sl Phys. Lett.} {\bf B238} (1990) 41\semi
D. Garfinkle, {\sl Phys. Rev.} {\bf D41} (1990) 1112\semi
D. Garfinkle and T. Vachaspati, {\sl Phys. Rev.} {\bf D42} (1990) 1960.}
\lref\garfink{D. Garfinkle, \prd{46}{1992}{4286}}
\lref\dan{D. Waldram, {\sl Phys. Rev.} {\bf D47} (1993) 2528.}
\lref\ghrw{J. Gauntlett, J. Harvey, M. Robinson, D. Waldram,
{\sl Nucl. Phys.} {\bf B411} (1994) 461.}
\lref\ghH{J. Gauntlett and  J. Harvey,
``S-Duality and the Spectrum of Magnetic Monopoles in Heterotic
String Theory", EFI-94-36, hep-th/9407111.}
\lref\ghsdyon{J. Gauntlett and  J. Harvey, ``$S$-Duality and the
Dyon Spectrum in $N=2$ SuperYang-Mills Theory",
CALT-68-2017/EFI-95-56, hep-th/9508156.}
\lref\sethi{S. Sethi, M. Stern, E. Zaslow, ``Monopole and Dyon Bound States
in $N=2$ Supersymmetric Yang-Mills Theories", HUTP-95-A031, hep-th/9508117.}
\lref\bstring{G. T. Horowitz and A. Strominger, \npb{360}{1991}{197}.}
\lref\calmp{C. G. Callan, J. M. Maldacena and A. W. Peet,
``Extremal Black Holes as Fundamental Strings", PUPT-1565, hep-th/9510134.}
\lref\worldb{C. G. Callan, J. A. Harvey and A. Strominger,
\npb{367}{1991}{60}.}
\lref\worlds{C. G. Callan, J. A. Harvey and A. Strominger,
\npb{359}{1991}{611}.}
\lref\andy{A. Strominger, \npb{451}{1995}{96}, hep-th/9504090.}
\lref\gertra{R. Geroch and J. Traschen, {\sl Phys. Rev} {\bf D36} (1987) 1017.}
\lref\nacu{S. Naculich, {\sl Nucl. Phys.} {\bf B296} (1988) 837.}
\lref\dhs{M. Dine, P. Huet and N. Seiberg,
{\sl Nucl. Phys.} {\bf B322} (1989) 301\semi
J. Dai, R. G. Leigh, J. Polchinski, {\sl Mod. Phys. Lett.} {\bf A4}
(1989) 2073.}
\lref\hull{C. M. Hull, \plb{357}{1995}{545}, hep-th/9506194.}
\lref\dabh{A. Dabholkar,
\plb{357}{1995}{307}, hep-th/9506160.}
\lref\schwarz{J. Schwarz, ``An SL(2,Z) Multiplet of Type IIB Superstrings,''
\hfill\break CALT-68-2013, hep-th/9508143.}
\lref\dred{J. Schwarz, ``Dilaton - Axion Symmetry",
CALT-68-1815 (1992), hep-th/9209125\semi
J. Maharana and J. Schwarz, \npb{390}{1993}{3}\semi
A. Sen, \npb{404}{11993}{109}.}
\lref\burinski{A. Ya.~Burinskii, ``Complex String as a Source of Kerr
Geometry,'' hep-th/9503094; ``Some Properties of the Kerr Solution
to Low Energy String Theory,'' IBRAE-95-08, hep-th/9504139.}
\lref\disk{W. Israel, \prd{2}{1970}{641}\semi
C. A. Lo{' p}ez, \prd{30}{1984}{313}\semi
A. Burinskii, \plb{216}{1989}{123}.}
\lref\BBS{K. Becker, M. Becker, and A. Strominger, ``Fivebranes,
Membranes and Non-Perturbative String Theory", NSF-ITP-95-62, hep-th/9507158.}
\lref\harrshep{B. Harrington and H. Shepard, \prd{17}{1978}{2122}.}
\lref\tseyone{A. A. Tseytlin, \plb{251}{1990}{530}.}
\lref\tseytlin{A. A. Tseytlin, ``On Singularities of Spherically Symmetric
Backgrounds in String Theory,'' Imperial/TP/94-95/54, hep-th/9509050.}
\lref\chsrev{C. G. Callan, J. A. Harvey and A. Strominger,
``Supersymmetric String
Solitons'' in String Theory and Quantum Gravity 1991: Proceedings of the
Trieste Spring School,
World Scientific, Singapore, 1991, hep-th/9112030.}
\lref\susskind{L. Susskind, ``Some Speculations about Black Hole Entropy
in String Theory,'' RU-93-44, hep-th/9309145.}
\lref\sussuglu{L. Susskind and J. Uglum,
\prd{50}{1994}{2700}, hep-th/9401070.}
\lref\vafa{C. Vafa, as referenced in \susskind.}
\lref\rohmwitt{R. Rohm and E. Witten,
{\sl Ann. Phys.}(NY) {\bf 170} (1986) 454.}

\Title{\vbox{\baselineskip12pt\hbox{CALT-68-2028}
\hbox{EFI-95-67}\hbox{UPR-681T}\hbox{hep-th/9511053}}}
{\vbox{\centerline{Strings as Solitons \& Black Holes as Strings
}}}
{
\baselineskip=12pt
%\bigskip
\centerline{Atish Dabholkar$^{1}$, Jerome P. Gauntlett$^{1}$,
Jeffrey A. Harvey$^{2}$ and Daniel Waldram$^{3}$}
\bigskip
\centerline{\sl $^1$Lauritsen Laboratory of
High Energy Physics}
\centerline{\sl California Institute of Technology}
\centerline{\sl Pasadena, CA 91125, USA}
\bigskip
\centerline{\sl $^2$Enrico Fermi Institute, University of Chicago}
\centerline{\sl 5640 Ellis Avenue, Chicago, IL 60637 USA}
\bigskip
\centerline{\sl $^3$Department of Physics, University of Pennsylvania}
\centerline{\sl Philadelphia, PA 19104-6396, USA}
\bigskip
\medskip
\centerline{\bf Abstract}
\bigskip
Supersymmetric closed string theories contain an infinite tower of
BPS-saturated, oscillating, macroscopic strings in the perturbative
spectrum. When these theories have dual formulations, this tower of
states must exist nonperturbatively as solitons in the dual theories.
We present a general class of exact solutions of low-energy supergravity
that corresponds to all these states. After dimensional reduction they
can be interpreted as supersymmetric black holes with a degeneracy
related to the degeneracy of the string states. {}For example, in four
dimensions we obtain a point-like solution which is asymptotic to a
stationary, rotating, electrically-charged black hole with Regge-bounded
angular momentum and with the usual ring-singularity replaced by a string
source. This further supports the idea that the entropy of supersymmetric
black holes can be understood in terms of counting of string states. We
also discuss some applications of these solutions to string duality.
}
\bigskip
\Date{November 1995}
%\bigskip
\baselineskip=20pt plus 2pt minus 2pt
%\draft

\vfill\eject\pageno1

\newsec{Introduction}

The concept of `strong-weak' coupling
duality has led to surprising new insights
into the dynamics of supersymmetric four-dimensional gauge theories.
Duality provides equivalent descriptions of the same physics in
terms of either the `electric' or the `magnetic' variables.
When one description is strongly-coupled and complicated, the dual
description can be weakly-coupled and simple.
Many nonperturbative phenomena  in one description
can then be understood perturbatively in the dual description.
One important piece of information for relating the two descriptions
is the spectrum of BPS-saturated states. These
states belong to a `short' representation of the supersymmetry
algebra, and their mass is proportional to their charge.
As a result, the semiclassical spectrum of BPS states is
often reliable even at strong coupling \wittoliv.
Analysis of the BPS spectrum has played a crucial role
in motivating duality and also in many subsequent developments.
{}For example, in the simplest context of
$N=4$ super Yang-Mills theory, fundamental particles of the magnetic
theory correspond to BPS-saturated, magnetic monopoles in the electric
theory \refs{\foursusy}.
In global $N=2$ theories the BPS-saturated states are important
for a qualitative understanding of several dynamical phenomena such as
confinement and chiral symmetry breaking \refs{\swa,\swb}.

There is now increasing evidence for a similar duality between various
string theories.
It is natural to expect that
the spectrum of BPS-saturated states will be equally important for a
dynamical understanding of these dualities. For example, in
$N=2$ string theories one can construct a generalized
Kac-Moody Lie superalgebra in terms of BPS states, which governs
the form of the perturbative prepotential \hm.

By analogy with field theory, one might hope to find
fundamental strings themselves as solitons in the dual string theory
\refs{\sensix, \harvstro, \dabh, \hull}.
Moreover, because strings are extended, oscillating objects,
one might expect a much richer spectrum of such solitons.
As usual in strong-weak coupling duality this is problematic
for non-BPS states, but in string theory, many of the fundamental
string states are BPS-saturated and therefore should exist as
BPS solitons in a dual description.
In particular, the perturbative
spectrum of closed string theories contains
stable, macroscopic-string states
\refs{\wittms, \dabhharv} that are BPS-saturated.
{}For concreteness let us consider heterotic string theory
on $R^9 \times S^1$ and take the radius of the $S^1$ to be large
compared to the string scale.
A macroscopic string is a winding state
that wraps around this circle. In order to obtain a BPS-saturated state
the right-moving oscillators must be in the ground state, but an arbitrary
number of left-moving oscillators can be excited subject only to the
mass-shell condition. We thus obtain an infinite tower of stable,
oscillating, macroscopic strings.
Even though we regard
these states as macroscopic, string-like objects, one can always take the
radius of the circle to be quite small.
Such a dimensional reduction gives rise to an infinite tower of point-like
states in  one less dimension. {}For example, various electrically charged
supersymmetric black holes and their magnetic duals can be viewed
in this manner.

In this paper we present solutions of the low-energy string equations of
motion that are in close correspondence with this entire tower of
BPS states.
We take as our starting point the fundamental straight-string solution
obtained in \dghr . We then generate several new solutions by using various
solution-generating techniques, dimensional reduction, and duality
transformations.  We are thus able to study a wide variety of string-like
and point-like solutions with either `electric' or `magnetic' charges
in a unified framework. The original straight-string solution is
BPS-saturated,
preserves half the supersymmetries, and has the right structure of
chiral fermionic zero modes expected for a Green-Schwarz superstring.
All these properties are inherited by the transformed solutions in a
simple way.

Another motivation for this work is the suggestion that the entropy of
black holes can be understood in terms of degeneracy of massive string
states \refs{\susskind, \sussuglu, \russsuss, \vafa, \senentrop}.
In particular,  one can consider a supersymmetric, nonrotating
black hole with given mass and charges
in toroidally compactified heterotic string theory in four
dimensions and compare its  entropy with the number of BPS states in the
spectrum with the same  quantum numbers.
The degeneracy of these states can be computed reliably
in string perturbation  theory because  their mass and charges are not
renormalized. Sen has shown that this counting agrees with the entropy
associated with the stretched horizon of the corresponding black holes
\senentrop.
The  solutions that we describe  here
provide an attractive classical picture of this  degeneracy
in terms of the oscillations of an underlying string.

The outline of this paper is as follows.
In section two, we first describe the infinite tower
of BPS states in the perturbative spectrum of closed strings.
We then obtain the most general charged,
oscillating-string solution that corresponds to these states.
We begin with the family of traveling-wave solutions
constructed in \garfink\ by applying the generating
transform of \vachetal\ to the straight-string solution.
We then apply the `charged-wave' generating transform of reference
\refs{\senetal, \senchsol}.
By appropriately choosing the form of the traveling wave
we obtain an asymptotically flat solution which
correctly matches onto a string source.
{}Furthermore, we show that a BPS-saturated
solution is obtained only when the traveling wave and the
current have the right chirality to be consistent with the corresponding
state in the string spectrum. The asymptotic parameters can be readily
identified with the quantum numbers of the tower of BPS string states.
We then discuss the dimensional reduction of these
solutions to obtain string solutions in lower dimensions
by using periodic arrays.
This clarifies the relation between string
solutions in different dimensions.

In section three, we obtain a family of point-like solutions in
four dimensions by dimensional reduction of  the
oscillating-string solution in five dimensions.
The resulting four-dimensional solution
is identical to the supersymmetric limit of
a stationary, electrically-charged, rotating black hole solution
obtained by Sen \senblack.
The asymptotic parameters of the black hole are
all related to the profile of oscillation of the underlying string
in a simple way.
Moreover, the angular momentum is Regge-bounded as expected
for a state in the string spectrum.
Near the core, however, the naked ring-singularity
of the  four-dimensional
stationary solution is replaced by the much milder singularity of a
five-dimensional string source.

In section four, we briefly review the string duality transformations
and discuss some properties of the infinite tower of
solitonic solutions obtained by dualizing the solutions of section two.
We point out that in four dimensions, by dualizing the supersymmetric black
holes of
section three one obtains the tower of rotating, magnetically-charged
states
that are required by S-duality \senfour.

Section five contains some concluding remarks.

\newsec{Macroscopic Fundamental Strings}

\subsec{Infinite tower of Macroscopic Strings}

We now describe the perturbative spectrum of
macroscopic strings on $R^9 \times S^1$.
The $S^1$ has radius $R$ which we take to be large
compared to the string scale and the
macroscopic string wraps around this circle.
The string states are then characterized by their winding number $n$ and
quantized momentum $m/R$ in the compact direction. Specifically,
if $p_R$ and $p_L$  are the right-moving and the left-moving momenta
respectively in the compact direction, we have
\eqn\momenta{p_R = (\frac{m}{2R} - {n R\over 2\apm}),
    \qquad p_L =(\frac{m}{2R} + {n R\over 2\apm}).}
It is convenient to use the Green-Schwarz formalism in the
light-cone gauge.
Let us first discuss the heterotic string which has $N=1$ spacetime
supersymmetry that is generated by right-moving worldsheet currents.
The supercharge has $16$ real components corresponding to
a single Majorana-Weyl fermion.
A BPS-saturated state preserves $8$ of these supersymmetries \dan ,
and the remaining $8$ broken supercharges generate
a $16$-dimensional `short' representation
of the $N=1$ superalgebra.
\foot{As discussed in \refs{\azcar, \dghr}, the $N=1$ supersymmetry algebra
in $D=9$ admits a central extension proportional to $p_R$.}

A `short' representation in spacetime can be obtained only
if all right-moving oscillators on the world-sheet are in the  ground state.
The left-movers can have arbitrary oscillations subject to the
Virasoro constraint:
\eqn\hetcon{
    N_L =1 + \apm\left(p_R^2- p_L^2-
    q_L^2\right) = 1-mn -\apm q_L^2,}
where $N_L$ is the number of transverse left-moving oscillators
and $q_L$ are the charges on the internal torus of the heterotic string.
Here we are working at a point in the Narain moduli space
where the lattice factorizes as $\Gamma_{0, 16} + \Gamma_{1, 1}$.
The mass of this state is related to the right-moving
momentum along the string
\eqn\massshell{ M^2 = 4 p_R^2.}

{}For type-II strings, we have $N=2$ supersymmetry
that is carried by both right-movers and left-movers.
Once again, a BPS-saturated state is obtained if at least
$8$ supersymmetries are unbroken which
can be either all right-moving or all left-moving.
Let us consider the spectrum when the unbroken supercharges
are all right-moving. In the worldsheet theory
these states are constructed as before by tensoring the right-moving
ground state with an arbitrary left-moving state subject
to the mass-shell conditions:
\eqn\twocon{\eqalign{
    N_L =&\, \apm\left(p_R^2 - p_L^2\right) = -mn,\cr
    M^2 =&\,  4 p_R^2.\cr }}
There is a corresponding tower of states when the
unbroken supersymmetries are all left-moving. All these
states preserve one quarter of the the original $32$
supersymmetries and belong to a `short' representation of the
supersymmetry algebra which is now $(16\times 256)$-dimensional.
A special case is when both right-moving and left-moving oscillators
are in the ground state. These states
preserve one half of the original supersymmetries and belong
to an `ultra-short' representation which is
$(16\times 16)$-dimensional.

Thus, the spectrum of BPS-saturated macroscopic strings
splits into topological sectors labeled by the winding
number $n$, and for every $n$ there is an infinite tower
of oscillating states with momentum $m/R$.
We denote these states by $(m , n)$.
Of course it is necessary to specify
the charge vectors
and which oscillators are excited
to completely label the state.
T-duality exchanges winding and momenta and in the case
of type-II theories takes the IIA theory to the
IIB theory \dhs.

{}For each theory there is an absolutely
stable macroscopic string state: $(1,1)$ for the heterotic theory and
$(0,1)$ for the type-II theories. The rest of the infinite tower of
BPS states are only neutrally stable. {}For example, the
multiply-wound state $(0,n)$ can decay into
$n$ singly-wound $(0,1)$ states sitting on top of each other.
However, because this tower of states exists already
in the single-string Hilbert space, they should be
regarded as distinct from multi-string states with
the same quantum numbers
\foot{For non-perturbative solitons this is not always the case.
{}For a discussion of related issues see, for example,
\refs{\swb,\ghsdyon,\sethi, \andy, \BBS, \wittD, \wittduality}.}.

This discussion can be readily generalized to compactifications
to lower dimensions.
The mass shell formula \massshell\ remains unchanged, but the
constraint formulae \hetcon\ and \twocon\ now receive
contributions from the internal conformal field theory.
The nature of these BPS states for compactifications of
the heterotic string theory with $N=4$ and $N=2$ supersymmetry
are described in \schsen\ and \hm\ respectively.

\subsec{Fundamental Straight Strings}

The massless bosonic fields that couple to a macroscopic string are
the dilaton $\phi$ , the metric tensor $g_{\m\n}$,
the antisymmetric tensor $B_{\m\n}$ and gauge field
$A_\m$. {}For simplicity we shall take the gauge group to be a single
$U(1)$.

In Type-II theories all these fields come from the
Neveu-Schwarz Neveu-Schwarz (NS-NS) sector
because, to lowest order, a fundamental
string does not couple to the massless fields from the
Ramond-Ramond (R-R) sector.
In ten dimensions there are no gauge fields in the NS-NS sector,
but after compactification they may be present in lower dimensions.
As we shall see in section four, some of the solitonic strings do
couple to R-R fields, but in that case, the R-R fields are related to
the NS-NS fields by a duality transformation, and need not be
considered separately. There can also be more general stringy solitons
that couple to both the NS-NS and the R-R fields
\refs{\harvstro , \hull , \schwarz}, however we do not
discuss them in this paper.

In heterotic string theory these fields are present
in all dimensions. In ten-dimensional heterotic string theory the
gauge group can be either $SO(32)$ or $E_8 \times E_8$, but we can
break it to $U(1)^{16}$ by turning on Wilson lines along the circle
that the macroscopic string wraps around.

The low-energy action for these fields in $D$ dimensions
\refs{\gsw} is given by
\eqn\action{S = {1 \over {2\kappa^2}}\int d^{D}x \sqrt{-g}
    e^{-2 \phi}\left(R + 4(\nabla \phi)^2 - {1 \over 12} H^2
    - {\apm } F^2 \right),}
where $F=dA$ and $H$ is
the three-form antisymmetric tensor field strength
\eqn\anomaly{H=dB -2\apm\omega^{YM}_3(A)
			+\ldots \, ,}
with $\omega_3$ the Chern-Simons three-form. Note that the
Lorentz Chern-Simons term will play no role in our solutions.

We also need to know the source terms
for various fields in the presence of a macroscopic string.
These couplings can be calculated directly by vertex-operator
calculations as in \wittms.
Equivalently, one can use the sigma model action as in \dghr :
\eqn\sig{\eqalign{
    S_\sigma &= -\frac{1}{4\pi\apm}\int d^2\sigma\left(
         {\sqrt\gamma}\gamma^{mn}\p_mX^\mu\p_n X^\nu g_{\mn}
         + \epsilon^{mn}\p_mX^\mu\p_n X^\nu B_{\mn} \right) + S_A, \cr
}}
where $X^\mu$ are the spacetime coordinates of the string,
and $\e_{mn}$ is the antisymmetric tensor-density on the
worldsheet with $\e_{01}=1$. The gauge field coupling
given by $S_A$ in \sig\
is somewhat delicate due to world-sheet anomalies in
the left-moving currents; this leads to an important quantum modification
of the current and source terms which will
be discussed in \S2.6.
At this order, there is no explicit source term for the dilaton,
but the dilaton couples to the string implicitly
through its coupling to other fields.

The equations of motion that follow from the combined action are
\eqn\emotion{\eqalign{
    &4\nb^2\phi - 4\left(\nb\phi\right)^2 + R - \frac{1}{12}H^2
          - \apm F^2 = 0 \cr
    &\nb_\mu\left(e^{-2\phi}H^{\mn\rho}\right) =
          \frac{\kappa^2}{\pi\apm\sqrt {-g}} \int d^2\sigma
          \epsilon^{mn}\p_mX^\n\p_nX^\r
          \delta^{(D)}\left(x-X(\sigma)\right) \cr
     &R^\mn + 2\nb^\mu\nb^\nu\phi
          - \frac{1}{4}H^{\mu\rho\sigma}{H^\nu}_{\rho\sigma}
          - 2\apm F^{\mu\rho}F^\nu_{\ \ \rho} \cr
          &\qquad \qquad \qquad =
          - \frac{\kappa^2\etp}{2\pi\apm\sqrt {-g}}
          \int d^2\sigma {\sqrt\gamma}\gamma^{mn}\p_mX^\mu\p_n X^\n
          \delta^{(D)}\left(x-X(\sigma)\right) \cr
   &{\nb_\mu}\left(e^{-2\phi}F^\mn\right)
          + \half e^{-2\phi}H^{\nu\rho\sigma}F_{\rho\sigma} \cr
                   &\qquad \qquad \qquad =
          \frac{\kappa^2}{4\pi\apm^2\sqrt {-g}}
%          \int d^2\sigma \epsilon^{mn}\p_mY\p_nX^\n
%          \delta^{(D)}\left(x-X(\sigma)\right). \cr
          \int d^2\sigma J^m\p_mX^\n
          \delta^{(D)}\left(x-X(\sigma)\right). \cr
                   &\qquad \qquad \qquad +
          \frac{\kappa^2}{2\pi\apm\sqrt {-g}}
          \int d^2\sigma \epsilon^{mn}\p_mX^\n\p_nX^\m A_\m(x)
          \delta^{(D)}\left(x-X(\sigma)\right). \cr
}}
After taking into account world-sheet anomalies, we shall show in \S2.6
that $J^m$ is the consistent anomaly current and that the full
gauge field source term (the right-hand side of the last equation)
becomes the covariant anomaly current.

The fundamental string solution \dghr\ of these equations
of motion is given by
\eqn\straight{\eqalign{
    ds^{2} &=- e^{2\phi} dudv +d\vx\cdot d\vx ,\cr
    B_{uv} &= \half ( e^{2\phi} -1),\cr
    e^{-2\phi} &= 1+{Q\over r^{D-4}} ,\cr
    A_{\m} &= 0, \cr}}
where $u=x^0 - x^{D-1}$ and $v=x^0 + x^{D-1}$ are the lightcone
coordinates along the string worldsheet,
$x^i$ (with $i= 1,...,(D-2)$) are the transverse
coordinates, $r$ is the radial distance in the transverse space,
and $x^{D-1} \equiv x^{D-1} + 2\pi R$.
By changing the sign of the axion field $B$, one can  reverse the
orientation  of the string.
The parameter  $Q$ is proportional to the ADM-mass per
unit length.  It was shown in \dghr\  that this solution
satisfies a Bogomol'nyi bound, admits a covariantly constant spinor
and preserves half the supersymmetries.
There is no momentum in the compact direction, and
the central charge in the supersymmetry algebra in $D-1$ dimensions
is proportional to the  winding number.
This solution can be identified with the field configuration
around a neutral macroscopic string state
$(0, n)$ by choosing
the ADM mass per unit length to be $n/(2\pi\apm)$.

\subsec{Neutral Oscillating Strings}

We would like to obtain a solution that carries charge as well as
arbitrary left-moving oscillations.
When the charge and the oscillations are small,
we could proceed by quantizing the collective coordinates
of the straight-string solution discussed in \S2.1;
this would be  analogous to constructing a dyon
with a small electric charge by quantizing the collective coordinates
around a neutral static magnetic monopole solution.
However, for large charge and oscillations, one would like
to directly construct a charged oscillating string-solution
analogous to the time-dependent Julia-Zee dyon solution in field theory.

In this subsection we discuss the neutral oscillating string.
To this end, we shall first review the solution-generating
transform of \vachetal . This transform was used in \garfink\ to construct
a general class of macroscopic string solutions with traveling waves.
However, as we shall see,
only a very small sub-class of these solutions
can be identified with the tower
of macroscopic string states that we are interested in.
To make this identification we require that the solution be
asymptotically flat and supersymmetric. We further require that it
match onto appropriate string sources. We apply these
criteria, in this and the following
subsections, to obtain the solution
that corresponds to an arbitrary $(m, n)$ state.

The solution generating transform of \vachetal\
can be applied to any solution of the equations
of motion \emotion\ which
admits a null Killing vector $K^\m$ that is hypersurface-orthogonal.
This means that the vector $K$ and the corresponding one-form satisfy
the conditions
\eqn\kform{
    K^2 =\, 0, \qquad L_K{ \Psi} =0, \qquad dK = K \wedge d M,}
where $\Psi$ refers to all fields, $L_K$ is the Lie-derivative
along $K$, and $M$ is a function that is determined by the solution.
The straight-string solution~\straight\ admits two such Killing vectors
$\partial/ \partial v$ and $\partial /\partial u$ with
$M= -2\phi$,
so the generating  technique can be used to construct waves that are
right-moving and left-moving respectively.
{}For a left-moving wave  the resulting solution is given by
\eqn\btrav{\eqalign{
    ds^2 =& - e^{2\phi}\left(dudv-T(v,\vx) dv^2 \right)
                   + d\vx\cdot d\vx \cr
    B_{uv} =& \half (e^{2\phi} - 1) \cr
    e^{-2\phi} =& 1+{Q\over r^{D-4}} \qquad \partial^2 T(v,\vx) = 0}}
where $\partial^2$ is the Laplacian operator in the flat transverse
space. It is easy to see that the equations of motion for $\phi$
and $B_{\m\n}$ are unchanged with the new fields. It was shown in
\vachetal\ that the Einstein equations are also unchanged
if $T$ is a solution of the flat Laplacian.
The wave travels at the speed of light because the solution still admits
$\p /\p u$ as a null Killing vector.
Replacing $T(v,\vx) dv^2$ with $T(u, \vx) du^2$ one obtains the
right-moving traveling-wave solutions.

The solution for $T(v,\vx)$ can be decomposed as a power series in
$r$ using $(D-2)$-dimensional spherical harmonics $Y_\ell$,
\eqn\spharm{
    T(v,\vx) = \sum_{\ell\geq0} \left[a_\ell(v)r^\ell
                    +b_\ell(v)r^{-D + 4 -\ell} \right] Y_\ell .}
Terms that go as $r^\b$ with $\beta =0$ can be removed by a change of
coordinates.
The remaining terms in this expansion have a direct physical interpretation.
Terms with $\beta\ge2$ and $\beta\le-D +3$ can be interpreted as
gravitational plane-fronted waves superposed on the string background.
Note that the former are not asymptotically flat and while the
latter are, they do not contribute to the asymptotic
quantities such as the ADM mass and, moreover, they
do not match onto string sources.
The most relevant term for our purpose is the one
with $\beta = 1$.
We shall see in $\S2.6$ that this is the only term that
matches onto a string source corresponding to an oscillating string.
The term  with $\beta= -D+4$ is also interesting even though
it does not match onto a classical string source;
asymptotically it corresponds to a
momentum wave without oscillations.
Keeping
only these two terms, the general form of $T$ is given by
\eqn\asympD{
    T(v,\vx) = {\vec f(v)}\cdot\vx + p(v)r^{-D+4}.}
In these coordinates, the solution does not appear asymptotically flat
because $T$ does not vanish at large $r$.
To remedy this we perform the following coordinate transformation:
\def\fdot{{\dot{\vec F}}}
\def\vf{\vec F}
\eqn\cotransf{\eqalign{
    v &= v' \cr
    u &= u' - 2{\fdot \cdot \vx '} + 2{\fdot\cdot\vf}
         - \int^{v'} \fds {dv} \cr
    \vx &= \vx ' - {\vf} \cr }}
where the dot indicates a derivative with respect to $v$,
${\vec f(v)}=-2{\ddot{\vec F}}$, and $\fds = \fdot\cdot\fdot$.
The metric and the other fields in the new coordinates are
\eqn\oscillating{\eqalign{
    ds^2 = -e^{2\phi}du\, dv +& \left[\etp\, p(v)\, r^{-D+4}
     -\left(\etp -1\right){\fds} \right] dv^2 \cr
        &+2 \left(\etp -1 \right){\fdot \cdot d\vx} dv + {d\vx\cdot d\vx}, \cr
      B_{uv} =& \half(\etp -1), \cr
      B_{vi} =& {\dot F}_i (\etp -1),  \cr
      e^{-2\phi} =& 1 + \frac{Q}{|\vx -\vf|^{D-4}},  \cr
}}
where we have suppressed the primes on the new variables.
The metric is now asymptotically flat and one can easily compute
a mini ADM 'stress-energy' tensor giving the mass and the momentum
per unit length, $\pi^{0,\mu}_{ADM}$, together with the momentum flow
along the string, $\pi^{D-1,\mu}_{ADM}$. We find
\eqn\adm{
    \left( \matrix{\pi^{0,\mu}_{ADM} \cr
          \pi^{D-1,\mu}_{ADM} \cr} \right)
    = k \left( \matrix{
          Q + Q \fds + p & Q{\dot F}^i& -Q \fds - p \cr
          -Q \fds - p & -Q{\dot F}^i& -Q + Q \fds + p \cr } \right),
}
where $k={(D-4) \omega_{D-3} \over 2\kappa^2}$
with $\w_{D-3}$ being the volume of the sphere $S^{D-3}$.

We shall see in \S2.6 that the neutral string solutions with $p=0$
map onto a
string source with a traveling wave with profile $\vec F$. Consequently,
they can be identified with the neutral states $(m,n)$
\foot{The state $(1, 1)$ in heterotic string theory is an exception
because it has positive momentum and no oscillations.
We would like to identify this state with
the solution with $\vec F =0$ and $p<0$,
as will be discussed in
\S2.6.}.
The ADM formulae are consistent with this
interpretation because the momentum is negative
and the mass is modified in accordance with the constraints
and mass-shell conditions.

As in \dghr\ there also exist multi-string solutions.
In general, we have a solution of the form \btrav\ provided
$\partial^2e^{-2\phi(\vx )}=0$  and $\partial^2T(v,\vx )=0$.
A general multi-string solution is given by,
\eqn\mult{\eqalign{
    e^{-2\phi(\vx)} &= 1 + \sum_{i} {Q \over
                        \left|\vx -\vx_i\right|^{D-4}}, \cr
    T(v,\vx) &= {\vec f(v)}\cdot\vx + \sum_{i} {p_i(v) \over
                        \left|\vx -\vx_i\right|^{D-4}}. \cr }}
%
%with $T(v,\vx)$ given by \asympD, with, for simplicity, $p=0$.
Making a coordinate transformation of the form \cotransf, we find
\eqn\multidil{
    e^{-2\phi} = 1 + \sum_{i} {Q \over
                        \left|\vx -\vx_i-{\vec F}(v)\right|^{D-4}} \, . }
Clearly, the constants $\vx_i$ determine the locations of the
strings, and we find that all the strings carry the same
profile of oscillation $\vf$, though in general carry different
momentum profiles $p_i(v)$.
These multi-string solutions will be useful later when we
discuss dimensional reduction using periodic arrays of strings.

\subsec{Charged Oscillating Strings}

Charged solutions can be obtained by applying  the solution generating
transform of \senchsol .  The basic idea behind this transform is as follows
\refs{\senetal, \veneziaetal}.
In string theory, the left-moving and the right-moving coordinates can be
rotated independently, except for the rotation of zero modes.
Consider string fields that are independent of
$p$ left-moving and $q$ right-moving compact coordinates of an
internal torus
as well as the time coordinate $t$.
Because the rotation of zero modes is unimportant for the coordinates
that the fields do not depend on
the string field equations of motion
and their low-energy limit are expected to possess
an $O(1, p)_L \times O(1, q)_R$
symmetry.
In the low-energy theory
these rotations correspond to nonlinear field transformations described
in \senetal .
Now, if a solution of the equations
of motion is not invariant under this symmetry, then one obtains a
new  `twisted' solution.  Some of the symmetries, corresponding
to rotations of the internal coordinates,
change the parameters of the
internal torus. Thus, if we want to keep the parameters of the internal
torus fixed, then the moduli space of solutions  is
$(O(1, p)/O(p)) \times (O(1, q)/O(q)) $.

The fundamental string solution is independent of time and also of the
internal coordinate corresponding to the $U(1)$ gauge group which we
take to be left-moving. So, we can twist \straight\ using
an $O(1,1)_L$ rotation to obtain  a one parameter family of
charged strings carrying left-moving current. It turns out that
the resulting solution \refs{\senchsol, \dan}\ admits
a null, hypersurface-orthogonal, Killing
vector $\partial / \partial v$. So we can apply the
the generating technique  of the previous subsection
to obtain the most general
solution representing a charged string with traveling waves:
\eqn\gen{\eqalign{
    ds^2 =& -e^{2\phi}\left(dudv-W(v,\vx) dv^2 \right) + d\vx\cdot d\vx \cr
    B_{uv} =& \half (e^{2\phi} -1 ) \cr
    A_v =& \frac{q(v)}{r^{D-4}+Q} \cr
    e^{-2\phi} =& 1+{Q\over r^{D-4}} \cr
    W =& { {\vec f(v)}\cdot\vx } +\frac{p(v)}{r^{D-4}}+
        \frac{2\apm q(v)^2}{Q \left(r^{D-4}+Q\right)}. \cr
}}
This solution represents a charged, superconducting
string with a left-moving electric current as well as
left-moving oscillations.
The current is equal in magnitude
to the charge per unit length and is  proportional  to $q$.
We have also generalized the charge to a charge wave $q(v)$. The
most general solution of this form has $A_v=N(v,\vx)\etp$ with
$\partial^2N(v,\vx)=0$, but we shall assume that the string is only a
monopole source and
drop the higher spherical harmonics in the full solution for $A_v$.

One can obtain an asymptotically flat  spacetime by a coordinate
transformation exactly as in the previous subsection.
The ADM stress-energy is then given by
\eqn\admch{
    \left( \matrix{\pi^{0,\mu}_{ADM} \cr
          \pi^{D-1,\mu}_{ADM} \cr} \right)
    = k \left( \matrix{
          Q + Q \fds + \frac{2\apm q^2}{Q} + p
              & -Q \fds - \frac{2\apm q^2}{Q} - p
              & Q{\dot F}^i \cr
          -Q \fds - \frac{2\apm q^2}{Q} - p
              & -Q + Q \fds + \frac{2\apm q^2}{Q} + p
              & -Q{\dot F}^i \cr } \right).
}

In heterotic string theory in ten dimensions
we expect macroscopic string solutions
with left-moving oscillations and only left-moving charges;
these are clearly included in our construction.
In toroidal compactifications
of the heterotic and type II strings we expect
solutions with both left-moving and right-moving charges.
We expect the most general solution can be obtained from
our solutions by
a combination of dimensional reduction and boosting. We leave
the details of this to future work.

\subsec{Supersymmetry}

We have already discussed that the BPS
states in the first-quantized
string spectrum are invariant under half or one quarter of the spacetime
supersymmetries. If the spacetime solutions we have described are to be
identified with such string states we must be able to show that they too
preserve the same fraction of the spacetime field supersymmetries.
We now show this by explicit calculation, restricting
to ten dimensions for simplicity.

Let us first consider the heterotic string.
In a given background of bosonic fields, the supersymmetry
transformation  laws  of the fermion fields are
\eqn\ssf{\eqalign{
\delta\lambda_{-}&=\left(\gamma^\mu\partial_\mu\phi-{1\over 12}H_{\mn\rho}
\gamma^{\mn\rho}
\right)\epsilon_+\cr
\delta\psi_{+\mu} &=\left(\partial_\mu+{1\over 4}{{\Omega_{-\mu}}^{ab}}
\Gamma_{ab} \right)\epsilon_+\cr
\delta\chi&=F_{\m\n}\gamma^{\m\n}\epsilon_+,\cr
}}
where $\lambda_-$ is the dilatino,
$\psi_{+\mu}$ is the gravitino, $\chi$ is the gaugino,
$\e_+$ is the supersymmetry parameter and we have introduced the
generalized spin connection $\Omega_{-\m}=\omega_\m-H_\m/2$.
All spinors are Majorana-Weyl and the subscript denotes their
chirality. Also $\gamma^\mu,\Gamma^a$ are gamma matrices with
spacetime and tangent space indices, respectively.
The general charged string  with left-moving oscillations
is given by \gen;
substituting
into \ssf\ we obtain
\eqn\ssfsubL{\eqalign{
    \delta\lambda_{-} &= -\hf \p_i\phi\Gamma^i\Gamma^-\Gamma^+\epsilon_+ \cr
    \delta\psi_{ +v }&= \left[ \partial_v - \frac{1}{8}e^\phi
        \left(\p_iW+2W\p_i\phi\right)\Gamma^i\Gamma^+ \right] \epsilon_+ \cr
    \delta\psi_{+u }&= \left[\partial_u
        + \frac{1}{4}e^\phi\p_i\phi\Gamma^i\Gamma^+ \right] \epsilon_+ \cr
    \delta\psi_{+i} &= \left[\partial_i - \frac{1}{2}\partial_i\phi
        - \frac{1}{4}\p_i\phi\Gamma^-\Gamma^+ \right] \epsilon_+ \cr
    \delta\chi&= e^{-\phi} \p_i A_u \Gamma^i\Gamma^+\epsilon_+ ,\cr
}}
where
$\Gamma^\pm=\Gamma^0\pm\Gamma^1$.
It is easy to see that the variations vanish when $\epsilon$
is of the form
\eqn\kspin{
    \epsilon_+ =e^{\phi/2}\epsilon_{+0}, \qquad \partial_\mu\epsilon_{+0}=0,
        \qquad \Gamma^+\epsilon_{+0}=0. }
\noindent Therefore, the solution admits a Killing spinor field of a
particular helicity along the worldsheet, and half the supersymmetries are
preserved.

Let us now consider a heterotic string with right-moving oscillations
and zero charge which is given by
\btrav\ with $T(v,\vec x) dv^2$ replaced with $T(u,\vec x) du^2$.
{}For this case
$\delta\lambda$ and $\delta\psi_i$ transform as before but
the remaining components transform slightly differently:
\eqn\ssfsubR{\eqalign{
    \delta\psi_{+v} &= \p_v\epsilon_+ \cr
    \delta\psi_{+u} &= \left[ \partial_u - \frac{1}{8}e^\phi
        \left(\p_iT\Gamma^i\Gamma^- - 2\p_i\phi\Gamma^i\Gamma^+\right)
        \right] \epsilon_+ .\cr }}
One can easily verify that there are no nonzero spinors for which
these variations vanish and hence all supersymmetries are broken.
So, in accordance with the states in the perturbative spectrum of the
heterotic string,
a BPS-saturated, partially
supersymmetric solution exists only if the oscillations are left-moving.

In the case of type-II strings we have two supersymmetry charges,
and the analysis for each of them parallels the discussion in
the preceding paragraphs. In IIA theory in ten dimensions,  if we set the
R-R fields to zero,
we can put the supersymmetry transformations in
the form
\eqn\ssfA{\eqalign{
    \delta\lambda &= \left[ \p_\mu\phi\gamma^\mu\Gamma_{11}
        +{1\over 6}H_{\mn\rho}\gamma^{\mn\rho} \right] \eta \cr
    \delta\psi_\mu &= \left[ \p_\mu + {1\over 4}\left(
        {\omega_\mu}^{ab}+{H_\mu}^{ab}\Gamma_{11}\right) \Gamma_{ab}
        \right] \eta , \cr}}
where $\Gamma_{11}=\Gamma^0\dots\Gamma^9$ and
the $d=10$ Majorana spinor $\eta$ can be decomposed into two Majorana-Weyl
spinors of opposite helicity, $\eta=\epsilon_++\epsilon_-$.
If we substitute the uncharged solutions with either left-moving
traveling waves, \btrav, or right-moving traveling waves,
\btrav\ with $T(v,\vec x) dv^2$ replaced by $T(u,\vec x) du^2$,
we again find Killing spinors
of the form
\eqn\kspinA{
    \epsilon_\pm=e^{\phi/2}\epsilon_{\pm 0}, \qquad
    \partial_\mu\epsilon_{\pm 0}=0 }
with the following helicity conditions. If $T=0$, then  there is
no traveling wave and both right-moving and left-moving oscillators
are in the ground state. In this case we have
\eqn\fundksA{
    \Gamma^-\epsilon_+ = 0 \qquad \Gamma^+\epsilon_- = 0, }
and half of the $N=2$ supersymmetries are preserved. If $T\neq 0$
then we have a traveling wave solution moving in either direction
and we obtain
\eqn\travksA{\eqalign{
    \hbox{left-moving }
        \Rightarrow\ &\epsilon_+ = 0,\quad \Gamma^+\epsilon_- = 0 ,\cr
    \hbox{right-moving }
        \Rightarrow\ &\Gamma^-\epsilon_+ = 0,\quad \epsilon_- = 0 .\cr }}
In each case only one quarter
of the $N=2$ supersymmetries are preserved.

A similar story unfolds for the type IIB string. Setting the
R-R
fields to zero, the supersymmetry transformations can be written as
\eqn\ssfA{\eqalign{
    \delta\lambda &= \left[ \p_\mu\phi\gamma^\mu \eta^*
        -{1\over 6}H_{\mn\rho}\gamma^{\mn\rho} \eta \right] \cr
    \delta\psi_\mu &= \left[ \p_\mu \eta
        + {1\over 4}{\omega_\mu}^{ab}\Gamma_{ab}\eta
        - {1\over 4}{H_\mu}^{ab} \Gamma_{ab} \eta^* \right] , \cr }}
where $\eta$ is a Weyl spinor and $\eta^*$ its complex conjugate.
Substituting the general form of the solution with left-moving or right-moving
traveling waves we obtain
Killing spinor fields of the form \kspinA, with the helicity conditions
\eqn\ksB{\eqalign{
    \hbox{fundamental string }
          \Rightarrow\ &\Gamma^+\epsilon^1_+ = 0,
\quad \Gamma^-\epsilon^2_+ = 0 \cr
    \hbox{left-moving }
          \Rightarrow\ &\Gamma^+\epsilon^1_+ = 0,
\quad \epsilon^2_+ = 0 \cr
    \hbox{right-moving }
          \Rightarrow\ &\epsilon^1_+ = 0,
\quad \Gamma^-\epsilon^2_+ = 0 \cr }}
where we have decomposed $\eta$ into its real and imaginary
Majorana-Weyl
components $\eta=\epsilon^1_+ +i\epsilon^2_+$.
Just as in the type IIA case the fundamental string solution preserves half
the supersymmetries, the traveling wave solutions, one quarter.

In summary, the supersymmetry of the spacetime solutions exactly
mirrors that of the corresponding BPS states in the
perturbative spectrum \foot{The matching of fermion zero modes
for Type II strings requires one to take into account a peculiar
chirality flip between static gauge and light-cone gauge which
is exactly analogous to that discussed in \worldb}.

\subsec{Sources}

In order to make a clear identification of our solutions with
the underlying fundamental string states we must show
that near the core
they match onto appropriate string sources which
satisfy both the string equations
of motion and the Virasoro constraints.
To do this in an unambiguous way we shall
take the radius of the  macroscopic string to be much
larger than both the string scale and the scale of
compactification.
For the neutral solutions this means we
can ignore the contribution
to the world-sheet stress tensor from the internal conformal
field theory and from normal ordering. Therefore, we need
to solve only the classical Virasoro constraints for
$D$ large dimensions. For charged strings on the other hand
we are forced to include one-loop anomaly effects in order
to get a consistent picture. This is forced by the close
connection between the Chern-Simons couplings in the low-energy
field theory and sigma-model anomalies in the string world-sheet
action \hullwitt.
We first discuss the source terms for the neutral
strings and then the charged strings.

Near the core, it is convenient to use the coordinates
used in \btrav\ even though they are not asymptotically flat.
On the worldsheet we fix conformal gauge and use
$\sigma^\pm=\tau\pm\sigma$.
{}For the spacetime fields we take the ansatz \btrav ,
and  for the string coordinates we take
\eqn\atzII{
    U = U(\sigma^+ ,\sigma^- ) \qquad V = V(\sigma^+ ,\sigma^- )
\qquad \vX = 0. }
The equations of motion for the string coordinates are
\eqn\streqn{\eqalign{
    g_\mn & \p_m\left({\sqrt\gamma}\gamma^{mn}\p_m X^\nu\right)
    + \Gamma_{\mn\rho}\left({\sqrt\gamma}\gamma^{mn}
       \p_m X^\nu\p_n X^\rho\right) \cr
    & + \hf H_{\mn\rho}\left(\epsilon^{mn}\p_m X^\nu\p_n X^\rho\right)
%    + F_{\mu\nu}  \left(\epsilon^{mn}\p_m Y\p_n X^\nu\right)
    = 0, \cr }}
where $\Gamma_{\mn\rho}$ is the Christoffel symbol for $g_\mn$.
With our ansatz, the $u$, $v$ and $i$ components become
\eqn\steqsub{\eqalign{
    \etp \left( \p_+\p_-U - 2T \p_+\p_-V
        - \p_v T \p_+V\p_-V \right) &= 0 \cr
    \etp \p_+\p_-V &= 0 \cr
    \p_i\left(\etp T\right) \p_+V\p_-V
           - \p_i\left(\etp\right)\p_+U\p_-V&=0\cr
}}
%        - \p_iA_v\left(\p_+Y\p_-V-\p_-Y\p_+V\right) &= 0 \cr }}
%
respectively. Similarly, the Virasoro constraints are given by
\eqn\ViraII{\eqalign{
    T_{++} &= -\etp\left( \p_+V\p_+U - T \p_+V\p_+V \right) = 0 \cr
    T_{--} &= -\etp\left( \p_-V\p_-U - T\p_-V\p_-V \right) = 0 \cr }}
{}Finally, the equations of motion for the spacetime fields reduce to
\eqn\subsrcII{\eqalign{
    & \partial^2e^{-2\phi} = -\frac{\kappa^2}{\pi\apm}
         \int d\sigma^+d\sigma^-
         \left( \p_+V\p_-U - \p_-V\p_+U \right)
         \delta^{D}(\vx - \vX ) \cr
    & \partial^2e^{-2\phi} = -\frac{\kappa^2}{\pi\apm}
         \int d\sigma^+d\sigma^-
         \left( \p_+V\p_-U + \p_-V\p_+U \right)
         \delta^{D}(\vx - \vX ) \cr
    & 0 = \int d\sigma^+d\sigma^-
         \p_+V\p_-V \delta^{D}(\vx - \vX ) \cr
    & T\partial^2e^{-2\phi} + e^{-2\phi}\partial^2 T
         = -\frac{\kappa^2}{\pi\apm}
         \int d\sigma^+d\sigma^-
         \partial_+U\partial_-U \delta^{D}(\vx - \vX ) \cr
}}
coming from the $vu$ component of the $H_{\mn\rho}$ equation, the $vu$,
$vv$ and $uu$ components of the Einstein equation,
respectively.

There is an important subtlety in solving these equations because
the sources are not ordinary functions but are Dirac
delta-functions in the transverse space. Do such
distributional sources lead to unique spacetime solutions?
One can regard the Dirac delta-function as a limit of
a smooth function. Now, there are many different smooth functions
which, in a limit,  represent a delta-function.
Geroch and Traschen \gertra\ show that in general,
due to the non-linear nature
of general relativity, the limiting
spacetimes are in fact different
depending on the function one uses to represent the delta function.
However, for metrics whose
curvature tensor
(or in our case the left-hand side of \subsrcII)
itself is a
distribution, the limiting spacetimes are the same.
Therefore, if we want a distributional
string source to uniquely determine a spacetime solution,
the left-hand sides of \subsrcII\ must also be distributions.

Another subtlety is that for all the solutions that we
consider, $\etp = 0$ at $r=0$. As a result,  \steqsub\
can be satisfied without constraining the string
coordinates at all provided the string is infinitely thin
and lies precisely at $r=0$. However,
if we replace the delta-function in \subsrcII\ with a smooth distribution
then the equations must be solved even slightly
away from $r=0$ where $\etp$ is nonzero. This is what
we shall do in the following.

We are now ready to determine which of the solutions described
earlier match onto sources. Let us first discuss
the straight-string solution with $T=0$.
In this case, the left-hand sides of \subsrcII\ are indeed distributions.
As shown in \dghr ,
we can satisfy all equations of motion and constraints if we choose
$V =V(\s^+ )$ and $U =U(\s^- )$.
Using the residual conformal invariance we can write the
solution in the form
\eqn\srcsol{
    U= 2 R n\sigma^- , \qquad V =  2 R n \sigma^+ .}
Here we have chosen the normalization such that
the string wraps $n$ times around the
circle of radius $R$ as $\s$ goes from $0$ to $\pi$.
The constant $Q$ in the spacetime solution is then
determined by the source to be
\eqn\Qhere{
Q={n\kappa^2\over \pi\alpha'(D-4)\omega_{D-3}},
}
which is such that the ADM-mass per unit
length is equal to $n$ times the string tension
$n/(2\pi\apm)$.

{}For the traveling wave solution \btrav,
we find that the left-hand side of the $uu$
component of the Einstein equation is not a distribution if $T$ diverges
at $r=0$. Hence neither the solutions
with $\beta\le -D+3$ (see the discussion after
\spharm), nor the solutions with nonzero $p$ ($\beta = -D+4$) can
be interpreted as arising from delta-function sources.
If we further require that the solution be asymptotically flat,
then $T$ is completely determined to be
${\vec f}(v)\cdot\vx$.
Using this form of $T$ the
string and source equations can be easily solved.
Because $T=0$ at the source we can solve the Virasoro
constraints \ViraII\ by taking
\eqn\srcsol{
    U= (2 R n +a)\sigma^- , \qquad V =  2 R n \sigma^+ .}
where $a\equiv {1\over \pi}\int^{2\pi Rn}_0 \dot{F}^2$ is
the zero mode of $\dot{F}^2$. We shall shortly
see that this is a convenient
choice. The constant $Q$ in the spacetime solution is again given by
\Qhere.
Thus, the oscillating string solutions are the only
asymptotically flat solutions that correctly match onto a string source.

As it stands, the solution \srcsol\ appears to correspond not
to an oscillating string at all, but
rather to a straight string lying at $r=0$. However,
if we transform to the primed coordinates \cotransf\
in which the metric is manifestly asymptotically flat, we obtain
\eqn\srctrnsf{\eqalign{
    V' &= 2 R n \sigma^+ \cr
    U' &= (2 R n +a)\sigma^-
      + \int^{V'}\dot F^2
        \cr
    \vX' &= {\vec F}(V'), \cr }}
which is indeed an oscillating string with profile ${\vec F}(V')$.
One can directly check that the Virasoro constraints are satisfied
using the metric \oscillating\ at the source.

The identification of this solution with the spacetime fields
around an oscillating string state can be further clarified by
taking $\kappa^2\to 0$
so that the string is very weakly
coupled to the spacetime fields.
This implies that
$e^{2\phi}\to 1$ and
uniquely defines a flat-space limit.
One can then directly verify that the string configuration \srctrnsf\
satisfies the Virasoro constraints in the limiting flat space and it is clearly
that of an oscillating string. The
momentum $p^\mu$ and winding vector $n^\mu$
of the string \srctrnsf\ in flat space can be
straightforwardly determined and we
find in the $({X'}^0,\vec X', {X'}^{D-1})$ coordinates
that $n^\mu=(0,\vec 0,n)$ and
$p^\mu=(2\apm)^{-1}
(2nR+a, \vec 0, -a)$, where we have used the fact that
$\vec F$ has no zero mode.
The momentum in the compact direction is defined as $m/R$ and, by
definition, the classical number oscillator is given by $N_L=-nm$. For the
string in flat space we thus have,
\eqn\note{
     m=-{Ra\over 2\apm},\qquad
     N_L={nRa\over 2\apm},\qquad
     a={1\over \pi}\int^{2\pi Rn}_0 dv \dot F^2 . }
Note that both the momentum in the compact direction and $N_L$ are
determined by the zero mode of $\dot F^2$, as one expects.

The ADM momentum of our full string solution, \adm,
is consistent with the momentum of the above string source in flat space.
This confirms that our spacetime solution really does correspond to that
of an oscillating string and that the momentum is not renormalized.
On this latter point lets us also note that the calculation of the winding
vector in flat space is equally valid for
the source \srctrnsf\ in the full solution. What about the momentum vector?
Starting from the string source in \sig, the momentum
conjugate to $X^\mu$ is given by:
\eqn\moment{
    p_\mu= \frac1{2\pi\apm}
              \int^\pi_0 d\sigma(g_{\m\n}\dot X^\n+B_{\m\n} X'^{\n}) . }
After substituting the quantities in \oscillating\
evaluated at the string source,
we find
a cancellation between various term and we are left with the momentum
vector in flat space.
Note that in flat space,
$p_\m$ comes from the $g_{\m\n}\dot X^\n$ term, while in the full solution
it comes from the $B_{\m\n} X'^{\n}$ term. This again reveals
that the momentum is not renormalized and is perhaps
analogous to the cancellation of perturbative contributions to
the energy density for straight string solutions originally
found in \dabhharv.

At this point we have provided substantial evidence that the
neutral traveling wave solutions with $p=0$ should be identified with
the neutral $(m, n)$ state. We have worked in the regime
where $N_L$ is large and
the radius of the circle is large compared to the string scale
so that the string source can be treated
classically. In this limit, an oscillating string solution can
be thought of as a coherent
superposition of large numbers of first-quantized states and should
not be sensitive to normal ordering effects such as the one in the Virasoro
constraint in \hetcon.

The solutions with $p\ne0, F=0$
have not played a role in our discussion so far
because they do not map onto classical source terms.
However, they are important for describing low-lying states.
{}For example, the state $(1,1)$ with $N_L=0$ in
the heterotic string theory has positive momentum and
no oscillations. Therefore, it cannot be identified with
a solution with $p=0, F\neq 0$ which asymptotically always has
negative momentum.
However, as advocated in \ghrw,
it can be identified with the string solution with $F=0$,
$p=-Q/2 R^2$ and $Q$ given by \Qhere.
Using the ADM mass formula \adm\ and the calculations in this
section, we see that asymptotically it corresponds to
a straight string with momentum $1/R$ in the compact direction and
mass $(2R-1/R)$.
Of course, for the low lying states, $\apm$ corrections could
be important and can modify the classical source.
Further evidence for this identification would be provided if one could show
that this solution can be mapped onto some kind of ``quantum string
source.''
Such a calculation was attempted in \ghrw\ but there are
subtleties in matching to the correct string states
which were overlooked, so this is still an open question.
Nevertheless,
the comparison of the classical scattering of these spacetime solutions
with scattering of $(1,1)$ states in string theory carried out
in \ghrw\ seems to support
this identification.

Now let us discuss how the string solutions with left-moving
charges and traveling waves
map onto source terms.
We expect that the charged strings can be mapped onto
sources coming from left-moving internal coordinates.

In order to do this correctly we must take into account
the relation between the spacetime Chern-Simons couplings
in the low-energy effective action and the
world-sheet chiral anomalies which arise for purely
left-moving currents. When we couple to external
gauge or gravitational fields the Chern-Simons terms
lead to a current inflow which is precisely balanced
by the anomalies of the world-sheet theory \inflow\ so
that overall charge and energy-momentum are conserved.
Because we include the Chern-Simons terms in our low-energy
effective action, we must also include one-loop world-sheet
anomalies in order to have a consistent $\alpha'$ expansion.

In discussing this we shall make use of the formalism
developed by Naculich \nacu. There it was shown
that at one loop the total gauge current on the world-sheet
is afflicted by the covariant
anomaly rather than  the consistent anomaly.
The point is that the Chern-Simons coupling in \action\
leads to an extra term in the gauge current (the last term in \emotion)
which is precisely
what is needed to turn the total gauge current into
the covariant anomaly current when
$J^m$ is the consistent anomaly current.

It was also shown in \nacu\ how to incorporate these quantum
effects in terms of classical chiral world-sheet bosons. This is ideal
for our needs since, as in the neutral case, we would like to
solve for both the spacetime and the worldsheet fields.
The last term in \sig\ is then given by
\eqn\can{
S_A=-{1 \over 4\pi\apm}\int d^2\sigma\left({\sqrt \gamma}\gamma^{mn}
    (\p_m Y-\sqrt{2\apm}A_m)(\p_n Y-\sqrt{2\apm}A_n)
    -\sqrt{2\apm}Y\epsilon^{mn}F_{mn}\right)
}
where $Y$ is a left-moving internal coordinate satisfying the constraint
\eqn\chcon{
\p_-Y=\sqrt{2\apm}A_- .}
and we have introduced the notation $A_m=\p_mX^\m A_\m$ and
$F_{mn}=\p_mX^\m \p_nX^\n F_{\m\n}$. Varying this action with respect to the
spacetime gauge field leads to \emotion\ where $J^m$ is given by
the consistent anomaly current
\eqn\Jcons{
    J^m = -\sqrt{2\apm}\sqrt{\gamma}\gamma^{mn}
            \left(\p_n Y - \sqrt{2\apm}A_n\right)
        -\sqrt{2\apm}\epsilon^{mn}\p_n Y .}

Note that the total action is gauge invariant provided $Y$ transforms
along with $A_\mu$ and $B_{\mu\nu}$, so that
\eqn\gaugesym{\eqalign{
    \delta A_\mu &= \nabla_\mu \Lambda \cr
    \delta B_{\mu\nu} &= 2\apm \Lambda F_{\mu\nu} \cr
    \delta Y &= \sqrt{2\apm} \Lambda .\cr }}
It is worth pointing out that the source terms \can\ are consistent
with dimensional reduction. If we reduce the
spacetime and source action with no gauge fields on a circle of
constant radius, then
we obtain a spacetime action with left-moving and right-moving gauge fields.
Using the formula presented in the next section one can show that the source
terms precisely give \can\ after setting the right-moving gauge fields
to zero. From this perspective, however, the constraint \chcon\ must be
imposed by hand.

The equations of motion for the string coordinates $X$ can be straightforwardly
derived and for our ansatz we are led to the first two equations
in \steqsub\ with $T$ replaced with $W$, while the $i$ component is
modified to give
\eqn\Xigauge{
    \p_i\left(\etp W\right) \p_+V\p_-V
        - \p_i\left(\etp\right)\p_+U\p_-V
        + 2\sqrt{2\apm}\p_iA_v\p_+Y\p_-V =0. }
The equation of motion for $Y$ is implied by the constraint \chcon.
The Virasoro constraints are given by
\eqn\gaugeVir{\eqalign{
    T_{++} &= -\etp\left( \p_+V\p_+U - W \p_+V\p_+V \right) +
         (\p_+Y-\sqrt{2\apm}A_+)^2 = 0 \cr
    T_{--} &= -\etp\left( \p_-V\p_-U - T\p_-V\p_-V \right) = 0. \cr }}
The equations of motion for the spacetime fields reduce to
the first three equations in \subsrcII\ and in addition:
\eqn\cse{\eqalign{
    W\partial^2e^{-2\phi} +& e^{-2\phi}\partial^2 W
         + 4\apm e^{-4\phi}\p_iA_v\p^iA_v \cr
         &= -{\kappa^2\over \pi\apm} \int
         d\sigma^+d\sigma^-\p_+U\p_-U \delta^{D}(\vx - \vX ) \cr
    A_v\p^2e^{-2\phi} -& \p^2\left(A_ve^{-2\phi}\right) \cr
         &= {\kappa^2\over4\pi\apm\sqrt{2\apm}} \int d\sigma^+d\sigma^-
         \left(\p_+Y-\sqrt{2\apm}A_v\p_+V\right)\p_-U
         \delta^{D}(\vx - \vX ) \cr
}}
coming from the $uu$ component of the Einstein equation and
the $u$ component of the $A_\mu$ equation, respectively.

We can now solve these equations for the general charged, traveling-wave
solution given in \gen\ except that we take $p=0$. The first three
equations in \subsrcII\ imply that $\p_-V=0$ and that the
constant $Q$ is given by \Qhere.
Whereas $T$
was zero at $r=0$ for the pure traveling wave solution, for the
general charged solution, $W$ at $r=0$ is $2\apm{q^2}/Q^2$. This means
that, while the second two terms on the right-hand side of the first
equation in \cse\ are zero distributions, the first term is no longer zero.
To solve this equation we must take
\eqn\pU{
    \p_+U = {2\apm q^2 \over Q^2} \p_+V . }
Finally, we must solve the second equation in \cse. Noting that $A_v=q/Q$ at
$r=0$, we find that the two terms on the left-hand side of the equation,
each a delta function in the transverse space, cancel against each other.
Solving for the right-hand side equal to zero, we get the further condition
\eqn\srccond{
    \p_+Y = {\sqrt{2\apm}q \over Q} \p_+V .}
Of course the value of the gauge field at $r=0$ can always be changed by a
gauge transformation. In general we find the condition \srccond\ is
$\p_+Y=\sqrt{2\apm}\left.A_v\right|_{r=0}\p_+V$, which is a gauge covariant
expression. We note that if the string is not wound around the internal $Y$
direction then we can always find a gauge transformation to put $Y=0$. This is
the analog of choosing coordinates where the traveling wave profile
is flat.

With the conditions \pU\ and \srccond\ we find that the string equations
of motion are also satisfied, except for the $v$ component of the $X^\mu$
equation which reads
\eqn\Xv{
    \etp\p_+\p_-U = 0. }
implying that $U=U_+(\sigma_+)+U_-(\sigma_-)$. In particular we note
that \srccond\ implies that there is no contribution from $Y$ to the Virasoro
constraints. From the left-moving constraint \chcon\ on $Y$ we find
that $\p_-Y=0$.

As before, using the residual conformal invariance, we can rewrite the
solution for the string coordinates in the convenient form
\eqn\chstr{\eqalign{
    U &= (2 R n +a_F + a_q)\sigma^-
             + {2\apm \over Q^2} \int^V q^2 \cr
    V &= 2 R n \sigma^+ \cr
    Y &= {\sqrt{2\apm} \over Q} \int^V q , \cr }}
where now
\eqn\defa{
    a_F \equiv {1\over \pi} \int_0^{2\pi Rn} {\dot F}^2
       \qquad
    a_q \equiv {1\over \pi} \int_0^{2\pi Rn} {2\apm \over Q^2}q^2 . }
Transforming to manifestly asymptotically flat coordinates this becomes
\eqn\asychstr{\eqalign{
    U' &= (2 R n +a_F+a_q)\sigma^-
              + \int^{V'} \left({\dot F}^2 + {2\apm \over Q^2}q^2 \right) \cr
    V' &= 2 R n \sigma^+ \cr
    {\vec X}' &= {\vec F}(V') \cr
    Y &= {\sqrt{2\apm} \over Q} \int^{V'} q \cr }}
which appears to describe an oscillating string with a profile
${\vec F}(V')$ in the transverse space and a profile $\int^{V'} q$ in
the internal space, which is manifest as a charged wave in the
external space. To pin down the state of the source string, we
again go to the weak coupling limit by letting
$\kappa^2\rightarrow0$. One
can then directly verify that the configuration \asychstr\ satisfies the
corresponding flat Virasoro constraints, and so we may
interpret the source as a
oscillating string state carrying a charge wave. Calculating the
momentum, $p^\mu$, and winding, $n^\mu$, vectors in
$({X'}^0,{\vec X}',{X'}^{D-1})$ coordinates gives
$n^{\mu}=(0,\vec 0,n)$ and
$p^\mu=(2\apm)^{-1}(2nR+a_F+a_q,\vec0,a_F+a_q)$.
The momentum in the internal $Y$  direction is $a_q$. Thus we
can identify the momentum in the ${X'}^{D-1}$ direction, the
classical oscillator number $N_L$ and internal charges $q_L^2$ as
\eqn\chmn{
    m = - {(a_F+a_q)R \over 2\apm}
        \qquad
    N_L = {(a_F+a_q)nR \over 2\apm}
        \qquad
    q_L^2 = - {a_q nR \over \apm}. }
Again the ADM momentum of the full solution \admch\ is consistent
with the momentum of the string source \asychstr\ in flat space.

So far we have discussed the solution only to the lowest
order in $\apm$, but it may be possible to extend these results
to higher orders. Away from the string source,
all solutions that we have described, essentially belong
to a general class of solutions called `chiral null models'
by Horowitz and Tseytlin \horotsey . These authors
have shown that there exists a particular
renormalization scheme in the worldsheet sigma-model in which
the lowest-order solution of a chiral null model
satisfies the string equations of motion
to all orders in $\apm$. Moreover,
Tseytlin has suggested \tseytlin\ that, even near the core,
the only effect of $\apm$ corrections will be to smooth
out the source on the scale of $\apm$. Tseytlin has
also proposed \tseyone\ an attractive and plausible picture of
how the sources might arise by resumming `thin handles' of the full
string partition function. These results suggest
that an exact conformal field theory should
exist corresponding to a macroscopic string.
Such a `conformal field theory' with sources would be
an interesting generalization of the usual notion
of an exact string solution.

\subsec{Dimensional Reduction}

Under toroidal compactification, the infinite  tower of
macroscopic strings reduces to various pointlike and string-like
BPS-saturated states in lower
dimensions. One can generate the solutions corresponding to
these states by dimensional reduction
of the macroscopic string solutions described earlier.

A macroscopic string solution can be reduced in two ways:
either in a direction transverse to the string to obtain a string-like
solution, or in the longitudinal direction to obtain a point-like
solution. We discuss the former in this subsection and the latter
in section three.
Dimensional reduction in a transverse direction
can be achieved by considering an infinite periodic array of strings.
This is similar to the way magnetic monopoles or H-monopoles
can be obtained from periodic arrays of instantons
\refs{\harrshep,\khuri,\ghl}.

{}For simplicity let us consider a macroscopic string
in six dimensions that wraps around the $x^5$ direction on a
large circle of radius R. We would like to reduce this solution along one
of the transverse directions, say $x^4$, on a circle of radius
$R_c$, $x^4\equiv x^4 +2\pi R_c$.
We can take $R_c$ to be much smaller than $R$ but still
much larger than the string scale.
To reduce the solution we consider an infinite periodic array of
oscillating strings
localized at $\vx =\vx_0 ,  x^4 = x^4_0 + 2\pi R_c k$ where
$\vx =(x^1, x^2, x^3)$ is a three-dimensional vector
and $k$ is an integer, with the same profile of oscillation $\vf(v)$.
This multi-string solution is determined by the dilaton \multidil\
\eqn\dilone{
    e^{-2\phi} = 1 + \sum_{k=-\infty}^{\infty}
                    {Q\over {r^2 + (x^4 -x^4_0 -2\pi R_c k)^2}} }
with $r^2 =|\vx -\vx_0-\vf(x)|^2$.
The infinite sum can be readily performed
\harrshep\ to obtain
\eqn\diltwo{
    e^{-2\phi} = 1 +  {Q\over 2 R_c r} {\rm \sinh}\,{\frac{r}{R_c}}
          ({\rm cosh}\,{\frac{r}{R_c}} - \cos{\frac{x^4-x^4_0}{R_c}} )^{-1} .}
{}For fixed $Q$, in the asymptotic limit $r \gg R_c$ or, equivalently,
averaging over $x^4$, \diltwo\ reduces to
\eqn\dilthree{
e^{-2\phi} = 1 +  {Q\over 2 R_c r}  + O(e^{-r/R_c}).}
We recognize this solution as the five-dimensional dilaton
given by \multidil\ with $D=5$ if we identify the five-dimensional
dilaton charge with $Q/{2R_c}$. We have thus constructed a solution
that at distances much larger than the scale of compactification
looks exactly like a five-dimensional oscillating string. On smaller scales,
however, it differs substantially.
In particular, near the core it dynamically
decompactifies and matches onto a six-dimensional source.

It is easy to generalize this construction to
an arbitrary toroidal compactification of an oscillating string.
We simply consider an array of strings on an arbitrary lattice
using the general multi-string solution \multidil . In general,
it is not possible to perform the infinite discrete sum analytically
but its asymptotic form can easily be extracted.

These considerations clarify the nature of sources in the previous
subsection. When we solved the constraint equations for a string
in $D$ large dimensions we ignored the contributions to the world-sheet
stress tensor from the internal conformal field theory. If the the
conformal field theory is a torus, then this is justified
when the size of the torus $R_c$  is much smaller than the radius
$R$ of the string. The solution then matches onto a $D$-dimensional
source. However, one can also consider intermediate scales
when $R$ and $R_c$ are both comparable and much larger than the
string scale. In this case, we effectively have a ten-dimensional theory
and the solution matches onto a ten-dimensional source.
Similar considerations should apply to a more general compactification.

We can also use dimensional reduction to understand
the relation between different types of string solutions. For
example,
consider a traveling wave solution with the
profile pointing in the compact direction. Then using the solution
in asymptotically flat coordinates and the techniques in the following
section we get precisely the charged string solution after
taking the limit $r>>R$. In particular the
charge wave is given by $q(v)=Q\dot F/{\sqrt 2}$.

\newsec{Black Holes as Strings}

We now turn to dimensional reduction of
macroscopic strings in the longitudinal direction to obtain
point-like ``black hole" solutions in one lower dimension.

In toroidal compactification of string theories, there
exists an infinite tower of BPS-saturated,
electrically-charged, point-like states in the perturbative
spectrum, which is closely related to the tower of macroscopic
strings discussed earlier. {}For example, in heterotic string theory
compactified to four dimensions
on a six-torus,  there is a tower of states
satisfying  the mass-shell conditions
\eqn\hetspectwo{\eqalign{
N_L =&\,  1 + \apm\left(q_R^2- q_L ^2\right), \qquad N_R = 0,\cr
M^2 =&\, 4 {q_R^2}.\cr
}}
Here $q_R$ and $q_L$ are the right-moving and the left-moving
charge-vectors respectively; together they define a point
in a $28$-dimensional, self-dual, Lorentzian lattice with
signature $(22, 6)$.
A typical such state has nonzero angular momentum $J$
which satisfies a Regge bound \gsw . The right-moving
ground state corresponds to a $N=4$ vector-multiplet in four dimensions
so can have maximum spin one. The maximum left-moving spin is $|N_L|$,
so the total angular momentum satisfies the Regge bound:
\eqn\regge{
J\leq |2+ \apm\left(q_R^2 - q_L ^2\right)|.}
Starting from these states,
S-duality predicts \senfour\ an infinite tower of magnetically charged
states, which must exist as solitons. We
shall discuss these solitons in section four.

Corresponding to the tower of first-quantized states
we expect to find BPS-saturated electric black-holes.
The most general, rotating,
charged, stationary black hole solution of the low-energy heterotic
string has been constructed by Sen \senblack .
We can try to identify this solution with the states discussed above
by choosing appropriate asymptotic quantum numbers, so that
in particular, the solution is supersymmetric.
However, we immediately run into a difficulty because to obtain
the supersymmetric limit one has to extend well
beyond extremality when the angular momentum is non-zero and the
solution has a severe ring-like naked singularity \burinski .
Moreover, the angular
momentum of these solutions can be arbitrary and is not
Regge-bounded.

The new solutions that we discuss in this section
are obtained by dimensionally reducing an oscillating string
in higher dimensions along the length of the string.
These solutions
are asymptotically identical to the supersymmetric, stationary,
rotating, electrically-charged, black-hole solutions
but near the core they have
a much milder singularity of a higher-dimensional string source.
Their angular momentum is naturally Regge-bounded because
the source solves the
string equations of motion and constraints.

In this section we restrict our attention to those states
in the spectrum whose entire left-moving and right-moving charge
arises from the winding and momentum along a single internal
circle. In this case, the charges $q_R$ and $q_L$ in
\hetspectwo\ are identical to the momenta $p_R$ and $p_L$
respectively in \hetcon . For simplicity, we shall also restrict
to the dimensional reduction of oscillating strings in five dimensions
to four.
These considerations can be easily extended
to more general cases.

\subsec{Dimensional Reduction to Four Dimensions}

The five-dimensional action is obtained by specializing
\action\ to $D=5$ and setting the gauge fields to zero:
\eqn\fiveaction{S = {1 \over { 2{\tilde \kappa}^2 }}\int d^{5}x
                     \sqrt{-{\tilde g}}
		    e^{-2 {\tilde \phi}}\left( R({\tilde g}) +
                      4(\nabla {\tilde \phi})^2
                                 - {1 \over 12} {\tilde H}^2 \right),}
where the tilde ${\tilde{}} $ is used to denote five-dimensional quantities.
{}For dimensional reduction, it is convenient to introduce \dred\
the `four-dimensional' fields
$g_{\m\n} , B_{\m\n} , \phi , A^a_{\m}$ and $\s$
$(0\leq \m \leq 3, a=1, 2)$ through the relations
\eqn\dimred{\eqalign{
e^{-4\s} \, =&\, {\tilde g}_{44} , \qquad\qquad\qquad\quad
                         \phi \, =\, {\tilde \phi} + \s , \cr
A^1_{\m} \, =&\, \half {\tilde g}_{4\mu} e^{4\s}, \qquad\qquad
A^2_{\m} \, =\, \half {\tilde B}_{4\mu}, \cr
g_{\m\n} \, =&\, {\tilde g}_{\m\n} - {\tilde g}_{4\m}
                   {\tilde g}_{4\n} e^{4\s},\cr
B_{\m\n} \, =&\, {\tilde B}_{\m\n} -2(A^1_{\m}A^2_{\n}
                   - A^1_{\n}A^2_{\m} ). \cr
}}
The low-energy lagrangian of a string compactified on a circle is
expected to have $SO(1,1)$ symmetry. This is broken to the perturbative
$SO(1,1,Z)$ T-duality symmetry by worldsheet instantons.
To make this symmetry manifest, we define
\eqn\ML{M = \left( \matrix{e^{4\s} & 0\cr
                           0       &  e^{-4\s}\cr} \right)
  \qquad\qquad L = \left( \matrix{0&1\cr 1&0\cr} \right)   }
which satisfy
\eqn\condition{ MLM^T =L, \qquad M^T =M}

To obtain the effective action in four dimensions, we express
the five-dimensional fields in \fiveaction\ in terms of
the four-dimensional fields  and take all fields to be independent
of $x^4$. The resulting action is
\eqn\fouraction{
\eqalign{S = {1 \over {2\kappa^2}}\int d^{4}x \sqrt{-g}
                e^{-2 \phi}&\left(R + 4(\nabla \phi)^2
                - {1 \over 12} H^2  - F^a (LML)_{ab}F^b\right)\cr
      &+\frac{1}{8} {\rm Tr} (\nabla M L \nabla M L ),\cr}}
where all spacetime indices are contracted with the four-dimensional
metric and $\kappa^2 \equiv \frac{{\tilde \kappa}^2}{2\pi R}$ is the
four-dimensional
gravitational coupling. The field strengths are defined by
\eqn\fstrength{F^a = dA^a, \qquad H = dB + 2 A^a \wedge F^b L_{ab}.}
Note that a Chern-Simons term appears in the definition
of $H$ in four dimensions even though it was absent in five dimensions.
{}For later use, it is more convenient to use `left-moving' and
`right-moving' gauge fields defined by
\eqn\lrgauge{ A^L = \frac{-A^1 + A^2}{\sqrt 2}, \qquad\qquad
A^R = \frac{A^1 + A^2}{\sqrt 2},}
so that the anomaly term for the two gauge fields is
diagonal.

By applying these reduction formulae to the neutral oscillating
string solution \oscillating\ with $p=0$ and reducing along the direction
$x^4$ (recall $u=x^0-x^4$, $v=x^0+x^4$) one obtains the four-dimensional
quantities:
\eqn\redsol{\eqalign{
e^{-4\s} \, =&\, \left( e^{2\tphi} - \fds (e^{2\tphi} -1) \right),\cr
e^{4\phi} \, =&\, -g_{00} = e^{4\tphi + 4\s},\cr
g_{ij} \, =&\, \d_{ij} - \fd_i\fd_j e^{4\s} (e^{2\tphi} -1)^2 ,\cr
g_{0i} \, =&\, \fd_i\fds e^{4\sigma} (e^{2\tphi} -1)^2
             + \fd_i (e^{2\tphi} -1) ,\cr
A^1_0 \, =&\, -\half \fds e^{4\s} (e^{2\tphi} -1), \qquad\qquad
              A^1_{i} \, =\, \half \fd_i e^{4\s} (e^{2\tphi} -1) , \cr
A^2_0 \, =&\, -\half (e^{2\tphi} -1), \qquad\qquad
              A^2_{i} \, =\, \half \fd_i (e^{2\tphi} -1), \cr
B_{0i}\, =&\, \fd_i (e^{2\tphi} -1) +\half \fd_i (\fds -1)
e^{4\s} (e^{2\tphi} -1)^2 ,\cr
}}
where
\eqn\fivedilaton{
e^{-2\tphi} = 1 +\frac{Q}{|\vx - \vf (v)|}.}
Note that this solution still has dependence on the internal coordinate.
To obtain a true four-dimensional solution one must do the usual
Kaluza-Klein truncation as we explain in the following. Before continuing
to analyze this solution we first review the stationary extremal
black hole solutions of \senblack. We shall see that the solution \redsol\
asymptotically approaches these solutions.

\subsec{Supersymmetric Stationary Black holes}

The most general stationary solution of \fouraction\ that
describes a rotating, electrically-charged black hole was constructed
by Sen \senblack . {}For our purpose we shall be interested in
only the supersymmetric solutions,
which in the notation of
\senblack, correspond to the limit
$\beta \rightarrow \infty , m \rightarrow 0$ keeping
$m_0=m \cosh (\beta )$ and $\a$ fixed.
The resulting solution depends upon three parameters,
$a$, $m_0$, and $\a$. Introducing
\eqn\bhdelta{
\D= (r^2 + a^2 \cos^2 \th )^2 +2m_0 \cosh \a (r^2 + a^2 \cos^2 \th ) +
      m_0^2 r^2 }
the solution can be written as
\eqn\bhsolution{\eqalign{
e^{4\phi}\, =&\, \quad {\D}^{-1} (r^2 + a^2 \cos^2 \th )^2 , \cr
ds^2 \, \equiv &\,\quad g_{\m\n} dx^{\m}dx^{\n} \cr
      \, =&\, - e^{4\phi} dt^2 + (r^2 + a^2 \cos^2 \th )
         ({r^2 +a^2})^{-1} dr^2 +
             (r^2 + a^2  \cos^2 \th ) \big\{d\th ^2 \cr
     &\, + \D^{-1} \sin^2 \th  \left[ \D +
    a^2 \sin^2 \th  (r^2 + a^2 \cos^2 \th  + 2 m_0 r \cosh \a)\right]
      d\varphi^2 \cr
       &\, - 2m_0 a r \sin^2 \th  \D dt d\varphi\big\} ,\cr
A^L_0 \, =&\, -\frac{m_0 n_L}{\sqrt 2} \D^{-1} r \sinh \a
         (r^2 + a^2 \cos^2 \th ) ,\cr
A^L_{\varphi} \, =&\,   - \frac{m_0^2 n_L}{\sqrt 2} \D^{-1} a \sinh
           \a r^2 \sin^2 \th (r^2 + a^2 \cos^2 \th ) , \cr
A^R_0 \, =&\, -\frac{m_0 n_R}{\sqrt 2} \D^{-1} r \cosh \a
      (r^2 + a^2 \cos^2 \th ) , \cr
A^R_{\varphi} \, =&\,   +\frac{m_0 n_R}{\sqrt 2} \D^{-1} a r \sin^2 \th
                  (r^2 + a^2 \cos^2 \th  +m_0 r \cosh \a ) , \cr
B_{0\varphi} \, =&\, - m_0 a \sin^2 \th r \D^{-1} (r^2 + a^2 \cos^2 \th  +
m_0 r \cosh \a ),}}
where $n_L$ and $n_R$ determine the sign of the charges and
can be either $\pm 1$.
{}For discussing the asymptotic behavior of these fields at large $r$,
it is simpler to use cartesian coordinates
$t, z, x^1 , x^2$. The axis of rotation
of the black hole is in the $z$ direction, $x^i (i=1,2)$ are the coordinates
in the transverse plane, and $r^2=z^2 + (x^i)^2$. The asymptotic
fields are then given by
\eqn\bhasymptotic{\eqalign{
e^{4\phi}\, \sim \, &\, 1 - \frac{2m_0 \cosh \a}{r} , \cr
ds^2 \, \sim \, &\, -(1 - \frac{2m_0 \cosh \a}{r} ) dt^2 + dz^2 + dx^i dx^i
     -\frac{2 m_0 a}{r} \e_{ij} x^i dx^j dt, \cr
A^L_0 \, \sim \, &\, -\frac{m_0 n_L \sinh \a}{\sqrt 2 r}, \qquad\qquad
A^L_i \, \sim \, 0, \cr
A^R_0 \, \sim \, &\, -\frac{m_0 n_R \cosh \a}{\sqrt 2 r}, \qquad\qquad
A^R_i \, \sim \, \frac {m_0 n_R a \e_{ij}x^j}{ \sqrt{2} r^3}, \cr
B_{0i} \, \sim \, &\, -\frac {m_0 a \e_{ij}x^j}{ r^3}, \cr
}}
where $\e_{ij}$ is an antisymmetric tensor with $\e_{12} =1$.
{}From these asymptotic formulae, one
can easily  read off the mass $M$, the angular momentum $J$,
the left-moving and right-moving  charges $Q_L$ and $Q_R$, and the
magnetic moments  $\mu_L$ and $\mu_R$ to obtain
\eqn\bhparameters{\eqalign{
M\, =&\, \half m_0 \cosh \a , \qquad\qquad\, \ J\, =\, \half m_0 a ,\cr
Q_L \, =&\, \frac{m_0 n_L }{\sqrt 2} \sinh \a , \qquad\qquad\,
Q_R \, =\, \frac{m_0 n_R}{\sqrt 2} \cosh \a ,\cr
\mu_L \, =&\, 0, \qquad\qquad\qquad\qquad  \mu_R \, = \,
\frac{m_0 n_R a}{\sqrt 2} .\cr
}}
Recall that for a black hole of mass $M$ and angular momentum $J$,
the canonical Einstein metric ${\hat g}_{\m\n}$ takes the asymptotic
form \mtw\
\eqn\einmetric{
d{\hat s}^2\, \equiv {\hat g}_{\m\n} dx^\m dx^\n
\sim\, -(1-\frac{2M}{r}) dt^2 + dz^2 + dx^i dx^i
-\frac{4J}{r^3} \e_{ij} x^i dx^j dt.}
In four dimensions, the string metric is related
to the Einstein metric by the relation
$g_{\m\n} = {\hat g}_{\m\n} e^{ 2\phi }$.

\subsec{Supersymmetric Black Holes as Strings}

To obtain a rotating black hole in four dimensions from our dimensionally
reduced string solutions, we choose
the profile of oscillation $\vf$ in \redsol\ to be of the form
\eqn\rotprof{\vf \, =\, A ( {\hat e_1} \cos
\omega t + {\hat e_2} \sin \omega t ), }
where ${\hat e_i}$ are unit vectors in the $i$th direction.
This corresponds to a string configuration that is
rotating in the $x^1 -x^2$ plane with amplitude $A$ and frequency $\omega$.
The period of oscillation $T$ can
be taken to be very small, and on time scales
much larger than $T$, only time-averaged quantities will be observable.
These time-averaged quantities can be compared with the parameters
of the stationary solution. There is an ambiguity in directly
time-averaging the fields themselves.
{}For example, the average of a product of two fields will not be the
same as the product of averages if there are correlations between
the two fields. Of course, all physical quantities such as geodesic
distance, forces, and in particular all asymptotic charges
will have a well defined time-average. Therefore, we consider
time-averages of only the asymptotic fields.

The time-averaging is equivalent to the usual dimensional
reduction. In order to reduce a solution
that depends on the internal coordinate $x^4$, one has to first expand
every field $\Psi$  in terms of the momentum modes:
$\Psi (x^4 ) = \sum_n \Psi_n e^{i n x^4/R}$. {}From
the lower-dimensional point of view, the modes $\Psi_n$ are massive
with masses of order $n/R$. Unless we probe the structure of the solution
on scales much shorter than $R$, we see only the zero momentum mode
$\Psi_0$ which is nothing but the average
of $\Psi$. Because our solution depends on $x^4$ only through the
combination $x^4 +t$, averaging over $x^4$ is the same as time-averaging.

The two basic time-averages are
\eqn\faverage{\eqalign{
\langle F_i \fd_j \rangle \, =&\,\frac{A^2\omega}{2} \e_{ij} ,\cr
\langle \fd_i \fd_j \rangle \, =&\,\frac{A^2\omega^2}{2} \d_{ij} ,\cr
}}
where we use $\langle A \rangle$ to denote time-average of the
quantity $A$
over a period of oscillation. Using these relations we obtain
\eqn\asrot{\eqalign{
\av{e^{4\phi}} \, \sim &\, -\av{g_{00}} \, \sim \,  1 -
                 \frac{Q}{r} (1 + A^2 \omega^2) , \cr
\av{g_{0i}} \, \sim &\,  -\frac{Q A^2 \omega}{2 r^3} \e_{ij} x^j ,\cr
\av{B_{0i}} \, \sim &\,  -\frac{Q A^2 \omega}{2 r^3} \e_{ij} x^j ,\cr
\av{A^1_0} \, \sim &\, \frac{Q A^2 \omega^2}{2r}\qquad\qquad
           \av{A^1_{i}} \, \sim \, -\frac{Q A^2 \omega}{4 r^3} \e_{ij}x^j, \cr
\av{A^2_0} \, \sim &\, \frac{Q}{2r}\qquad\qquad\qquad
        \av{A^2_{i}} \, \sim \, -\frac{Q A^2 \omega}{4 r^3} \e_{ij}x^j .\cr}}
One can easily read off various asymptotic quantities:
\eqn\asrot{\eqalign{
M\, =&\, \frac{Q( 1+ A^2 \omega^2)}{4}, \qquad\qquad\,
\ J\, =\, \frac{Q A^2 \omega }{4},\cr
Q_L \, =&\, -\frac{Q}{2{\sqrt 2}} (A^2 \omega^2 - 1), \qquad\qquad\,
Q_R \, =\, -\frac{Q}{2{\sqrt 2}} ( 1 + A^2 \omega^2 ),\cr
\mu_L \, =&\, 0, \qquad\qquad\qquad\qquad \qquad  \mu_R \, = \,
-{\sqrt 2} J .\cr
}}
These are exactly the same parameters as in \bhparameters\
with $n_L =n_R = -1$.
The three parameters $m_0 ,\a$, and $a$ can be easily computed in terms
of $Q , A$, and $\omega$. The left-moving and right-moving
gyromagnetic ratios are defined by the relations
$\mu_L = \frac{g_L Q_L J}{2 M}$ and
$\mu_R = \frac{g_R Q_R J}{2 M}$. One easily obtains
\eqn\gyro{g_L =0, \qquad g_R = 2}
These gyromagnetic ratios are related to the left-moving
and right-moving angular momenta by the relations \refs{\russsuss ,\senblack}
\eqn\gyrotwo{
g_L = 2 \frac{J_R}{J_R +J_L} , \qquad g_R = 2 \frac{J_L}{J_R +J_L}.}
Thus, from these asymptotic quantities we find
that the angular momentum of our solution is carried
entirely by left-movers consistent with the fact that the
underlying string solution has only left-moving oscillations.
It is also satisfying that the right-moving gyromagnetic
ratio is two as expected for a fundamental object.

The discussion in the preceding paragraphs can be easily
extended to nonrotating charged black holes by
considering a slightly different profile of oscillation $\vf$:
\eqn\nonrot{\vf (t) = A {\hat e_1} \cos \omega t .}
The time-averages are somewhat different from \faverage
\eqn\faveragetwo{\eqalign{
\langle F_i \fd_j \rangle \, =&\, 0 ,\cr
\langle \fd_i \fd_j \rangle \, =&\,\frac{A^2\omega^2}{2} \d_{i1}\d_{j1} ,\cr
}}
The asymptotic parameters are then given by
\eqn\asnonrot{\eqalign{
M\, =&\, \frac{Q( 2+ A^2 \omega^2)}{8}, \qquad\qquad\,
\quad J\, =\, 0,\cr
Q_L \, =&\, -\frac{Q}{4{\sqrt 2}} ( A^2 \omega^2 -2), \qquad\qquad\,
Q_R \, =\, -\frac{Q}{4{\sqrt 2}} ( A^2 \omega^2 + 2),\cr
}}
which again agrees with  \bhasymptotic\ if we set $a=0$.
More generally, one can consider an arbitrary {}Fourier expansion
for $\vf$.

As a consistency check we note that, both for the rotating and
the nonrotating solution, the asymptotic form of the antisymmetric
tensor is also in agreement with the stationary solution.
In four dimensions, the antisymmetric tensor is equivalent to
an axion field $b$ through the relation $db = * dH$.
Now,  because of the anomaly term $ dH \sim {\vec E} \cdot {\vec B}$,
there is a source term for the axion whenever there is
a non-vanishing  ${\vec E} \cdot {\vec B}$.
A rotating charged object has both charge and magnetic moment and
accordingly a nonzero ${\vec E} \cdot {\vec B}$ that falls off
as $1/r^5$. One then expects a nonzero value of the axion
that falls off as $1/r^3$. In the non-rotating case, on
the other hand, one expects a faster fall-off. This is precisely the
behavior we find for the time-averaged fields in the two cases.

We now compare these results with the perturbative spectrum \hetspectwo .
{}For this purpose we need to properly
take into account different normalizations that we have used.
The normalization of charges $Q_L$
and $Q_R$ depends on the normalization of the gauge fields,
and is different from the normalization $q_L$ and $q_R$.
Furthermore, in the black-hole formulae above, we have
chosen units in which the four-dimensional Newton's constant
$G\equiv \kappa^2/ (8\pi)$ is one.
{}From the normalization of the sigma model action and the dimensional
reduction formulae, we readily see that $Q_{L,R}= 2 \sqrt{2} q_{L,R}$,
and from \Qhere\ that $Q= 4nR/{\apm}$.
If we consider a string that wraps $n$ times around the internal circle
of radius $R$ then the periodicity of string coordinates
implies that $\omega = \ell /nR$ for some integer $\ell$.

Using these formulae it is easy to see that various asymptotic
parameters of our black hole are in complete agreement with
the quantum numbers of a perturbative state in the spectrum
\hetspectwo .  In particular, $M=2q_R$ and the angular
momentum is given by
\eqn\ang{
J= \frac {A^2 \ell}{\apm}.}
The maximum angular momentum on the other hand is given
by \regge\ which from \asrot\ equals $A^2 \ell^2/ \apm$.
We thus see that the angular momentum is Regge-bounded
and the leading Regge trajectory corresponds to taking
$l=1$ for arbitrary $A$. This is in precise agreement
with the behavior expected from a string source at the core.

We would now like to comment upon the singularities
of the reduced string solution vis-a-vis those of the stationary
Sen solution. Recall, that the stationary Sen solution is obtained by
twisting the Kerr solution in Einstein gravity.
One of the peculiarities of the Kerr solution is the existence
of the Kerr singular ring \burinski\
which is a branch-cut connecting spacetime
on  `negative' and `positive' sheets. Across the branch-cut various
fields change their signs and directions.
A similar naked singularity exists also for the Sen solution;
across the branch-cut the dilaton becomes imaginary.
By contrast, although the reduced string solution asymptotically
approaches the Sen solution after time-averaging,
the solution has
only a mild singularity of a string source from the
five-dimensional point of view. In four dimensions,
the string source with a rotating profile is squashed to a
disk and it  may appear to have  a disk-like naked singularity.

We note that this is reminiscent of the proposal of \disk\
in which the negative sheet of the Kerr solutions is
replaced  by a disk-like source bounded by the Kerr ring.
It was argued in \disk\ that the disk must be super-conducting
and  in rigid rotation, and  in \burinski\ that
the boundary of the disk could be identified with some string-like source.

It is also worth comparing our solutions with those in \hands\
where rotating charged black holes were constructed in dimensions
$D>4$ with one non-zero component of the angular momentum.
It was found that for these dimensions
the BPS limit coincides with the extremal limit. By taking
periodic
arrays of these solutions in six dimensions, a four-dimensional
rotating black hole was constructed
by dimensional reduction that asymptotically
approached the four-dimensional stationary charged rotating black hole
solution. {}For these solutions the ring singularity of the four-
dimensional black hole is replaced by a six-dimensional BPS charged
rotating extremal black hole
singularity. One difficulty in relating these solutions to the
four-dimensional BPS states of string theory is that, unlike our
solutions, the angular momentum is not Regge bounded.

We conclude this section with some comments concerning the dimensional
reduction of
solutions with  $F=0, p\ne0$.
Starting with the five-dimensional solution with $F=0,p\ne0$
we obtain four-dimensional static supersymmetric black hole solutions with
mass and charge given by
\eqn\bdg{
M={Q+p\over 4},\qquad Q_L={1\over 2\sqrt 2}(Q-p),\qquad Q_R=
{1\over 2\sqrt 2}(Q+p).
}
For $p>0$ this static black hole solution is precisely the same as the
asymptotic behavior of the black hole solution that arises from the profile
\faveragetwo. This is to be expected because
even before dimensional
reduction, the solution with $p>0$ is asymptotically the same as the
time-averaged, non-rotating oscillating string solution. This
can be seen, for example, from the ADM tensor \adm.

It is also illuminating to consider the behavior of these static black hole
solutions under the $T$-duality transformations of
Buscher \busch. In our notation
these transformations take the form
\eqn\tdual{
\eqalign{
\sigma'&=-\sigma, \qquad \phi'=\phi+\sigma ,\cr
{A^1}'&=A^2,\qquad {A^2}'=A^1 ,\cr
g_{\m\n}'&=g_{\m\n},\qquad B_{\m\n}'=B_{\m\n} .\cr}
}
One finds that this has the effect of simply interchanging the
parameters $Q$ and $p$ in the solution. Now let us consider a
neutral $(m,n)$ state.
We have seen in the last section
that this corresponds to the neutral, oscillating-string solution
with $p=0$, $Q$ given by \Qhere, and a profile $F$
satisfying \note. Comparing how $F$ and $p$
contribute to the ADM momentum we deduce that asymptotically
the state $(m,n)$ corresponds to the static black hole with parameters
given by
\eqn\blah{
\eqalign{
Q&={n\over 2\pi\apm}{\tilde\kappa^2\over 2\pi}\cr
p&=-{m\over 2\pi R^2}{\tilde\kappa^2\over 2\pi}\cr
}}
Replacing the five-dimensional gravitational coupling $\tilde\kappa^2$
with $2\pi R\kappa^2$ and recalling that the four-dimensional
coupling $\kappa^2$ is held fixed under $T$-duality, we see that
the interchange of $p$ and $Q$ coming from spacetime duality
corresponds to the interchange of $n$ and $-m$ along with
$R$ and $\apm/R$ which are the worldsheet $T$-duality transformations.

It is also interesting to consider the solution with
$p<0$. From \bdg\ we see that we have constructed massless
black holes when we choose $p=-Q$.
In fact this construction is essentially the same as that in
\bekal.
Note that for the value of $p$ corresponding to
the heterotic string
state $(1,1)$, $p=-QR^2/\apm$, the black holes are massless
precisely at the self dual radius $R^2=\apm$.

\newsec{Strings as Solitons}

At present there are essentially three independent exact
string-string dualities which we list below.

\noindent I. The Type-IIA string compactified on a $K3$ surface
is conjectured to be dual to the heterotic string
compactified on a torus $T^4$.

\noindent II. The Type-IIB string in ten dimensions is conjectured
to be dual to itself.

\noindent III. In ten dimensions, the Type-I string and the heterotic string
with gauge group $SO(32)$ are conjectured to be dual to each other.

\noindent For our present purpose we shall be interested in only a
few general aspects of these dualities. A more
detailed description can be found, for example,
in \refs{\hulltown ,\wittduality}.

If two string theories are exactly dual to each other at
a microscopic level, the spectrum of BPS-saturated states
in the two theories must match. Under duality, perturbative
states in one theory typically match onto nonperturbative states
in the other. In this section we discuss the nonperturbative
states that are dual to the infinite tower of states analyzed
in previous sections. We then describe a variety of string-like and
point-like solitons in the low energy theory that correspond to
these states.

\subsec{Infinite Tower of Solitonic Strings}

Note that all string theories in the above list, with the exception
of the Type-I theory, are supersymmetric theories of closed strings only.
Consequently, they contain an infinite tower of macroscopic strings
in their perturbative spectrum.  Which states do these map onto under
duality? We address this question below.

A basic requirement of exact duality is that the low-energy action
for the massless fields in the two theories must agree after
appropriate field-redefinitions. One feature common to all dualities
is that strong coupling must match onto weak coupling, so the
dilaton $\phi$ maps onto $-\phi$. Furthermore, the Einstein-Hilbert
action should be left invariant, so
${\hat g_{\m\n}} \rightarrow {\hat g_{\m\n}}$ where
the canonical metric ${\hat g}$ is related to
the string metric $g$
by ${\hat g_{\m\n}} = e^{-{4\phi}/{(D-2)}}g_{\m\n}$
in $D$ `large' dimensions.
Transformations of the remaining fields such as the gauge
fields and the antisymmetric fields are specific to each case, and
are often nonlocal, but they
can all be deduced uniquely from the low-energy actions.

Given the duality map for the massless modes,
and their coupling to a fundamental macroscopic string, one can
immediately see  that the infinite tower of macroscopic strings
in each case must map onto a  nonperturbative state in the
corresponding dual theory.
A macroscopic fundamental string couples universally and
locally to the second-rank, antisymmetric, Kalb-Ramond field
$B_{\m\n}$ through the action \sig .
So the duality transformation of this field is particularly
important. For example, in the case of duality I above,
the $B_{\m\n}$ fields of the two theories are related
by a nonlocal transformation involving the field-strengths
$H$, $H \rightarrow *H$. Thus, the state that is dual
to a macroscopic string couples non-locally to the dual
antisymmetric tensor field. Such a state must arise
nonperturbatively because all perturbative states
have local couplings with this field. Similarly,
in the case of dualities II and III, the $B_{\m\n}$ field
maps onto a field coming from the Ramond-Ramond sector.
Once again there are no perturbative states that couple to the
Ramond-Ramond
field through the action \sig , so these must also
be nonperturbative.

It is trivial to find the solitonic solutions of the low-energy
equations of motion that correspond to this infinite tower of states
once we know the
duality map for the massless modes.
A solitonic solution is obtained
simply  by rewriting a solution in the previous
sections in terms of the new dual variables.

Even though the two solutions are related by a field redefinition,
their interpretation and properties are completely different.
{}First, the mass per unit length of a solitonic string in
a given theory is inversely proportional to the coupling constant
of that theory. So these states are infinitely massive at weak
coupling, as they should be.
Second, a fundamental string solution discussed
in previous sections has a source at the core;
it therefore describes the fields around a perturbative
macroscopic string state.
One cannot really regard them as solitons any more than one
can regard the fields around a classical electron as a soliton
in quantum electrodynamics. A solitonic solution, on the other hand,
has no sources.
Third, for a fundamental string, the coupling constant vanishes
near the core whereas for a solitonic string it diverges.
Consequently, the structure of singularities of the
string metric is quite different in the two cases.
{}Finally, the nature of higher order $\apm$ corrections
to the solutions also differs substantially in the two cases.

To illustrate these points let us consider the infinite tower
of solitonic strings required to exist by duality
in Type-IIA theory compactified on a $K3$ surface.
The mass of these states goes as $1/{\lambda^2}$ where
$\lambda^2$ is the loop counting parameter.
These are the solitons that are dual to the infinite
tower of perturbative states in the heterotic string theory
compactified on $T^4$. The solution corresponding to an infinitely
long straight-string soliton was analyzed  in \refs{\sensix ,\harvstro}.
At a generic point in the moduli space of $K3$, where the gauge symmetry
is $U(1)^{24}$, it was shown in \refs{\sensix ,\harvstro} that the
charged zero modes in the background of such a string live
on an even, self-dual, Lorentzian lattice with signature $(20, 4)$.
This is precisely the
structure expected for a charged heterotic soliton
with $N_L =0$. We would now like
consider a more general soliton corresponding to the state
$(m, n)$ at finite radius with nonzero oscillations.

We start with the general uncharged solution \btrav\ corresponding
to an oscillating,  heterotic macroscopic string in six dimensions
and rewrite it in terms of Type-IIA variables. The resulting solution
is given by,
\eqn\hetsol{\eqalign{
    ds^2 =& - \left(dudv- {\vec f(v)}\cdot\vx dv^2 \right)
                   +e^{2\phi} d\vx\cdot d\vx \cr
    H_{ijk} =&  -\e_{ijk}^{\quad\, l} \nabla_l{\phi}\cr
    e^{2\phi} =& 1+{Q\over r^{D-4}}, }}
where $i, j,  k, l$ are the indices in the space transverse to
the string, $\e$ is the completely antisymmetric tensor,
and $f$ is an arbitrary function as before.
We have used the coordinate frame that is not
asymptotically flat because we wish to concentrate
on the properties of the solution near the core.
Notice that near the core, ${\vec x}\rightarrow 0$, the solution
reduces to the one with no oscillations \ie\ the straight-string
solution discussed in \refs{\sensix , \harvstro}. This
has two important consequences.
{}First, the entire infinite tower of solitons is
smooth near the core and has an infinite throat just like
the straight-string soliton.
Second, near the core they will be described by
an exact superconformal field theory \refs{\worlds , \chsrev}
that consists  of a product of a WZW model and a free
field with a linear dilaton, with an additional perturbation
corresponding to the $dv^2$ term in the metric that becomes smaller
and smaller down the throat. Of course, because of the growing dilaton,
the theory eventually gets into strong coupling and the semi-classical
approximation breaks down.

If we consider the duality III, then we expect to find heterotic
winding states as solitons in Type-I theory. Because the mass of these
solitons goes as $1/{\lambda}$ they are very different
from the usual solitons in field theory. These solitons are also
a source of the Kalb-Ramond field in
the R-R sector of the Type-I theory,
and correspond to Dirichlet one-branes \refs{\brane,\braneone}.
The straight-string soliton was constructed in \refs{\dabh, \hull}.
For weak coupling and low winding number, interactions
are small, and this soliton
has a description in terms of an exact conformal
field theory of a one-brane \braneone.
It would be interesting
to generalize these results to oscillating one-branes.

\subsec{Infinite tower of Magnetically charged States}

The three basic string-string dualities, I-III, discussed above
imply other dualities in lower dimensions after
toroidal compactification. Under this reduction,
a solitonic string gives rise
to different  point-like and string-like solitons,
which is what we describe now. Our main
interest will be the magnetically charged states predicted
by S-duality in four dimensions.

Let us consider the heterotic string compactified
on $T^4 \times T^2$ which is dual to Type-IIA
on $K3 \times T^2$. It has been shown that the six-dimensional
string-string duality implies that each of the four-dimensional
theories are $S$-dual \refs{\wittduality,\duffss}.
In both theories there is
an infinite tower of electrically charged point-like
states given by  a formula analogous to \hetspectwo .
{}Four-dimensional S-duality then predicts an infinite
tower of magnetically charged states. We would now like
to show that the solutions corresponding to many of
these states can be obtained directly from solitonic
strings in six dimensions.

{}For example, consider the Type-IIA string which
has a heterotic string soliton in six dimensions.
We now consider a two step reduction of the soliton
to four dimensions. We first reduce in the $x^5$ direction,
as in $\S2.7$, by taking a periodic
array of solitons, to obtain a five-dimensional string soliton.
We then reduce in the $x^4$ direction, as in
$\S3.3$, to obtain a point-like solution in four dimensions.
The original string soliton has a `magnetic' charge
$\int_{\Sig}H$ where $H$ is the field strength of $B_{\m\n}$
and $\Sig$ is an asymptotic three-surface on the spatial slice
transverse to the string. As a result, the dimensionally
reduced soliton is magnetically charged under the gauge field
coming from $B_{4\m}$.  This is precisely the state
that is S-dual to a perturbative state that is electrically charged
under the gauge field coming from $g_{4\m}$.
Starting with an oscillating string
soliton in six dimensions, one can obtain an infinite tower
of magnetically charged states.
The degeneracy
of these states is naturally related to the degeneracy of
the oscillating string soliton.
Moreover, because these solitons are obtained from the
smooth string soliton, they are smooth near the core and have
an infinite throat much like the stationary solution \ghs.
The low lying states in this tower are particularly interesting.
For example, the lowest level of this
tower corresponds to magnetic monopoles \hl, the first
level to H-monopoles \refs{\rohmwitt, \khuri,\ghl} and so on.
Most of the analysis in this paper is applicable to states
with large oscillation number. It would be interesting to extend
it to these low-lying states.

{}This discussion makes it clear that the problem
of finding H-monopoles in four-dimensional heterotic
string theory is closely related to the problem
of finding the Type-IIA solitonic string moving on $K3$ in
six-dimensional heterotic string theory. It is difficult to see how
one can obtain translational zero modes for a soliton
that parametrize a complicated surface like $K3$.
The main difficulty is that such a motion is not associated
with any charges at infinity as is the case with the
motion on a torus. In particular, the charged zero modes
which are present for the heterotic string soliton in Type IIA
theory can be inferred using an anomaly inflow argument
directly from the low-energy theory \bhd. This is not true
for the IIA string as a soliton in the heterotic theory because
the zero mode structure after reduction to six dimensions
is completely non-chiral. These arguments seem to suggest
that the detailed
zero mode structure of the Type-IIA soliton cannot be
understood entirely within the massless theory. It would follow
that obtaining the precise degeneracy of H-monopoles also requires
a treatment beyond field theory, something that was also
apparent in \ghH\
for somewhat different reasons.

We do not have much to say about this important
problem except to point out the following possibility.
Consider a Type-IIA theory compactified on a K3 that is
geometrically an orbifold obtained from flat
space by modding out with a discrete group $G$. One
can consider a ten-dimensional  macroscopic string
that is located at a point other than the fixed points of G.
Its image under $G$ generates
a `periodic' array  of strings.  On distances larger
than the characteristic size of the $K3$, this periodic array
will describe a macroscopic Type-IIA string moving in
six dimensions. The zero modes that move it along the $K3$
will be visible, however, only on scales smaller than the
size of the $K3$. By taking the duality transform of this solution,
one can obtain, at least at large distances and away from
the orbifold singularities, the soliton
in heterotic string theory that corresponds to the Type-IIA string.

\newsec{Conclusions}

In this paper we have shown that there is a very satisfying agreement
between the structure of BPS states in toroidal compactifications
of supersymmetric strings and supersymmetric solutions to
the low-energy effective theory when source
terms are included. This leaves little doubt that
the solutions we have discussed are just the fields outside
of elementary string states.

This point of view has in the past led to a puzzle
when considering BPS fundamental string states carrying
angular momentum. In string theory the states have Regge-bounded
angular momentum and one expects source terms for the external
fields. On the other hand to reach the supersymmetric limit
for the known rotating black hole solutions one must go
beyond the extremal limit to very singular configurations, and
there is no reason for the angular momentum to be bounded. We think
we have resolved this puzzle by showing that the rotating BPS
string configurations are stable but depend on both time and
the internal coordinate. This leads
to a substantial modification of the short distance structure
of the solution which is consistent with an underlying
string source.  Upon time-averaging the solution, which is equivalent to the
usual Kaluza-Klein reduction, one obtains
the asymptotic form of the previously studied solutions. This
point of view may have broader applicability to black hole
physics in that it shows that information about black
hole structure may be carried by configurations which vary in
time on the string scale and which are averaged out in
a low-energy description of black hole physics.

There are a number of directions related to this work which
would be interesting to pursue. It would be very interesting
to extend this analysis to non-extremal black holes.
Both the uncharged and the charged
fundamental-string solutions are the extremal limit
of a larger class of black-string solutions representing string-like
singularities surrounded by event horizons \bstring .
Because the charged black-string solutions do not possess a null Killing
vector, it is not possible to use the generating technique to construct
traveling-wave solutions on these backgrounds.
Such a construction remains an open problem.

It would also be interesting to
see whether some of the problems of black hole information
loss can be resolved using a picture in terms of microscopic
string states as has been advocated by Susskind and others
\refs{\susskind, \sussuglu, \russsuss}.
A related question is the statistical interpretation of
black hole entropy.
The entropy of supersymmetric black holes as considered in
\senentrop\ now appears to have
a nice state-counting interpretation in terms of oscillations
of an underlying string.
This entropy has so far been computed
only for non-rotating black holes \senentrop. To make this picture
complete, one would like to generalize this computation to rotating
black holes.
We expect that the solutions discussed in this paper will
be useful for this purpose.
One puzzle is that in Sen's calculation the relevant
scale for determining the location of the stretched horizon
is the string scale and not the scale of compactification.
On the other hand, we have seen in this paper that the four-dimensional black
hole solution matches not onto a four-dimensional point source but rather
onto a five-dimensional string source.
Thus, beyond the scale of compactification one should
really use a five-dimensional description.
We hope to return to these questions in the near future.

We have also provided more evidence for the idea
that fundamental strings are solitons in the dual theory.
We have shown that the entire spectrum of macroscopic BPS states
exists as solitons.
This makes
it more plausible that even small, oscillating
loops of fundamental strings, which in general are unstable,
have a dual description.
There are many open questions in string duality
related to BPS states and macroscopic strings.
We have sketched a few possible applications
in section four; it would be nice to make these ideas more concrete.
We have so far explored the detailed structure
of these states only in theories with maximal supersymmetry.
Many new dynamical features emerge in theories with reduced
supersymmetry, so it would be useful to explore
the structure of BPS string states and their duals
in these theories as well.

It has also become clear recently that extended objects,
$p$-branes, play a fundamental role in duality.
We have seen that for strings, which are one-branes,
BPS states arise as purely left-moving excitations of the
underlying string. This suggests, at least for odd $p$,
that there may be a connection between excited BPS states
of these $p$-branes and ``chiral brane waves.''
This would be particularly interesting for
D-branes \refs{\brane, \braneone, \wittD, \branetwo}.

\bigskip

\leftline{ \bf Acknowledgements}
\bigskip
We wish to thank
D.~Garfinkle, R.~Geroch,  P.~Ginsparg, E. Martinec, M.~O'Loughlin,
A.~Sen, S.~Shenker,
and A.~Tseytlin for
valuable discussions.
A.~D., J.~P.~G., and J.~A.~H.
would like to thank the Aspen Center for Physics where some of
this work was completed.
This work was supported in part by the U.~S.~Department of Energy
under Grant No. DE-FG03-92-ER40701 and Grant No. DOE-AC02-76-ERO-3071,
by NSF Grant No. PHY 91-23780. and by NATO Grant No. CRG 940784.

\bigskip
\leftline{\bf Note Added} After this work was completed
we became aware of a paper with some overlap with ours \calmp.
\vfill
\eject

\listrefs
\end